\documentclass[12pt]{article}
\usepackage{epsfig}
\usepackage{amsmath}
\begin{document}

\begin{titlepage}
\begin{center}

Draft: \today	
{}~             
{}~		

\vskip1.0in 
{\Large\bf Quantum kinetics and thermalization in an exactly solvable model}

\vskip.3in

S. M. Alamoudi$^{(a)}$\footnote{Email: smast15@vms.cis.pitt.edu},
D. Boyanovsky$^{(a)}$\footnote{Email: boyan@vms.cis.pitt.edu},
H. J. de Vega$^{(c)}$\footnote{Email:devega@lpthe.jussieu.fr} and
R. Holman$^{(b)}$\footnote{Email: holman@defoe.phys.cmu.edu}\\ 

\bigskip

{\it (a) Dept.\ of Physics and Astronomy, University of Pittsburgh, 
         Pittsburgh PA USA 15260}\\
{\it (b) Department of Physics, Carnegie Mellon University, Pittsburgh,
         PA. 15213, U. S. A.}\\
{\it (c)  LPTHE, \footnote{Laboratoire Associ\'e aui CNRS, UMR 7589.}
Universit\'e Pierre et Marie Curie (Paris VI) 
et Denis Diderot  (Paris VII), Tour 16, 1er. \'etage, 4, Place Jussieu
75252 Paris, Cedex 05, France}

\vskip.3in

\end{center}

\vskip.5in

\begin{abstract}

We study the dynamics of relaxation and thermalization in an
exactly solvable model with the goal of 
understanding the effects of off-shell processes. The focus is to
compare the exact evolution of the distribution function with
different approximations to the relaxational 
dynamics: Boltzmann, non-Markovian and Markovian quantum kinetics.

The time evolution of the 
occupation number or distribution function is evaluated exactly using
two methods: time evolution  
of an initially prepared density matrix and by solving the Heisenberg
equations of motion. The former allows to  establish a connection with
the stochastic nature of thermalization and the 
fluctuation dissipation theorem, whereas the latter leads to the
interpretation of an interpolating number operator 
to `count' quasiparticles. There are  two different cases that are
studied in detail:  
i) no stable particle states below threshold of the bath and a
quasiparticle resonance  above it 
 and ii) a stable discrete  exact `particle' state below threshold.

The exact solution for the evolution allows us to investigate the
concept of the formation time of a quasiparticle and to study the difference
between the relaxation of  the distribution of particles and quasiparticles. 
In particular we compare the quasiparticle distribution for asymptotic
times with the equilibrium canonical distribution. For the case
of quasiparticles in the continuum (resonances) the
exact   quasiparticle  distribution asymptotically tends to
a  statistical  equilibrium distribution that  differs from a simple Bose-Einstein form as
a result of off-shell processes such as the strength of the
quasiparticle poles, the width of the unstable particle and proximity
to thresholds. In the case ii), the distribution of particles does
not thermalize with the bath. 
 
 We study the kinetics of thermalization and relaxation by deriving 
 a non-Markovian quantum kinetic equation which resums the
perturbative series and includes off-shell effects. A Markovian
approximation  that includes off-shell contributions and  
the usual Boltzmann equation are obtained from the quantum kinetic
equation in the limit of wide separation of time 
scales upon different coarse-graining assumptions.
 The relaxational dynamics predicted by the non-Markovian, Markovian
and Boltzmann approximations are  compared to the exact
result of the model. The Boltzmann approach is seen to  
fail   in the case  of wide
resonances and when threshold and renormalization  
effects are important. Implications for thermalization in field theory
models are discussed.   
\end{abstract}

\end{titlepage}

\setcounter{footnote}{0}


\section{\bf Introduction and Motivation}
The new generation of ultrarelativistic heavy ion colliders, RHIC at
Brookhaven and LHC at CERN will hopefully probe the 
quark-gluon plasma (QGP) and chiral phase transitions within the next few
years. These colliders will provide the first opportunity 
to study phase transitions predicted by QCD in controlled
experiments. The main goal of the theoretical 
program associated with ultrarelativistic heavy ion collisions is to
provide a thorough understanding of the experimental 
signatures associated with the formation and evolution of the QGP as
well as the chiral phase transition\cite{book1}. A very
important part of this program is to study the dynamical evolution
from the initial state 
corresponding to the highly contracted nuclei through the
thermalization stage and the onset of the QGP in local thermodynamic
equilibrium. The initial state after the collision is strongly out of
equilibrium and there 
are very few quantitative models to study its subsequent
evolution. Current theoretical predictions for the dynamics of the
initial 
 stages of formation of a QGP are based on the hard  and semihard
aspects which are studied via perturbative parton cascade
models. These assume that at large energy densities the nuclei can be
resolved into their partonic constituents and the dynamical evolution
can therefore be tracked by following the parton distribution
functions through 
a relativistic transport equation which includes scattering 
through the perturbative parton-parton interactions which are  dressed
by medium effects\cite{wang}-\cite{shuryak}.   

The important ingredient in this program is a relativistic kinetic or
Boltzmann equation for the phase space distribution function of
partons (quarks and gluons) whose 
collision terms incorporate the perturbative parton cross sections
including screening effects that cut-off the 
 infrared behavior\cite{eskola3,biro}.  Although  relativistic kinetic
transport equations  
provide a natural framework to study non-equilibrium phenomena and are
based on successes in many different fields (mostly 
at the non-relativistic level), their validity in the extreme
situations envisaged to arise during the early stages of 
heavy ion collisions at RHIC and LHC is questionable at best. Detailed
microscopic derivations of relativistic transport
equations\cite{elze1,geiger,dani,bass} reveal that several
coarse-graining assumptions must be invoked to arrive at the kinetic 
equations beginning from the microscopic Schwinger-Dyson
equations. The typical assumptions involve smooth and slow 
variations over microscopic time scales, which results in a condition
that the time scales for relaxation be much 
longer than the typical microscopic scales. These assumptions had been
discussed by Geiger\cite{geiger} within 
the context of parton cascade models of the evolution and
thermalization and more recently by Bass et. al.\cite{bass} within the
framework of the relativistic quantum molecular dynamics
approach. With time scales for thermalization predicted 
to be of order of $\approx 1 \mbox{fm}/c$ (chemical equilibrium is
typically achieved on longer time scales) 
the regime of validity of coarse-grained relativistic Boltzmann
equations  may be restricted to hard partons with typical
momenta $\geq \mbox{few Gev}$. However for soft quarks or gluons or
collective modes with typical scales determined by the  
Debye (or magnetic) screening lengths (at $T \approx 300-500
\mbox{Mev}$) $(m_D)^{-1} \approx 1/gT \approx 1 \mbox{fm}$  or wide
resonances, such as vector mesons, whether this separation of scales
is enough to justify the validity of the usual kinetic approach is
questionable.

 There is a
rekindled interest on a deeper understanding of quantum kinetics from
microscopic quantum field theory, rather than from a
semi-phenomenological approach and critical analysis 
of transport and kinetic approaches are beginning to
emerge\cite{makhlin}-\cite{morako}.  
In particular recently the initial stages of pre-equilibration during
which quasiparticle correlations begin to 
build had been investigated in a many body system\cite{morako}. The
pre-equilibrium stage cannot be studied within 
a Boltzmann approach because the early time dynamics depends on the
initial preparation of the state and is determined 
by off-shell (non-energy conserving) processes\cite{morako}.

Unlike non-relativistic cases in which the validity of kinetic
equations can be tested either numerically or experimentally as in
condensed matter physics, the reliability of the approximations
involved in obtaining relativistic kinetic equations 
in the high energy and density regimes to be probed at RHIC-LHC can
only be speculated at present. Therefore, it is important to obtain  
insight into the dynamics of thermalization and relaxation and the
validity of a kinetic description in simple model theories so as to
provide a yardstick to use as a guiding principle.  

It is the goal of this article to study the description of
thermalization and relaxational dynamics in a simple and 
{\em exactly solvable} many body theory to obtain a deeper
understanding of the off-shell processes involved in thermalization
and to provide a yardstick to test different approximations.  

The aspects that we seek to study  in this article are the following:
\begin{itemize}

\item{How do off-shell effects modify the dynamics of thermalization and
relaxation? By off-shell we here refer to processes that do not conserve
energy on short time scales and threshold effects that are not
incorporated in the usual Boltzmann equation. These are responsible
for quasiparticle 
properties such as widths and residues in the quasiparticle propagators.}

\item{ A detailed understanding of the relaxation of
quasiparticles vs. that of  bare and dressed particles and to explore
the definition of a quasiparticle 
distribution function that is valid beyond the narrow 
width approximation.}

\item{ A comparison of the validity of Markovian (coarse grained)
approximations including the Boltzmann equation,  to the most general
non-Markovian description of relaxation.}  
\end{itemize} 

Although we anticipate that the answer to these questions will in general
depend on the details of the microscopic model, we propose to study  a
model  which bears many properties of realistic quantum field theories
and allows us to provide the answers to these (and other) questions to
build intuition into more complex situations.  

In section \ref{secmodel} we introduce the model and analyze the
dynamics from the 
point of view of the time evolution of an initially prepared density
matrix and the solution of the Heisenberg equations of motion. This 
analysis allows us to establish contact with the fluctuation dissipation
relation. In section \ref{secgt} we study in detail the time evolution of the
distribution function and distinguish between bare particle and
dressed particle  and quasi-particle distributions. In section \ref{secnorm} 
we discuss the exact solution in  
terms of normal modes and analyze the definition of the interpolating
number operator that describes the relaxational dynamics. In section
\ref{seckin} we analyze the {\em approximate} relaxational dynamics in terms of
i) the Boltzmann equation, ii) the non-Markovian quantum kinetic
equation and iii) a Markovian approximation to the quantum kinetic
equation. 
In section \ref{secnum} we provide a numerical comparison of the exact and
approximate kinetics. Our conclusions and extrapolations to realistic
quantum field theories are summarized in section \ref{seccon}.

\section{\bf The Model \label{secmodel}}

As stated in the introduction, we seek to study aspects of quantum
kinetics in a model that allows to compare 
approximate treatments of the relaxational dynamics to exact
solutions. Obviously even a simple interacting  
quantum field theory would not allow  an exact treatment of the
dynamics and the validity of the approximations  
leading to the kinetic description could not be tested. Instead we
sacrifice generality and study  a model that 
has already been used in the literature to study aspects of
decoherence and dissipation phenomena, and which shares many 
of the important features of a quantum field theory, such as
renormalization and off-shell phenomena.

The physical situation that we have in mind is that of an initial
distribution of bare particles  that become in contact with a
medium of high energy density which we will take to be in equilibrium
at a given temperature $T$ and will be denoted by the `bath'.
Before interacting with the medium the particle(s) has a physical mass
and travels freely, which in field theory would correspond 
to a description in terms of physical {\em in} fields of the physical
mass. These bare particles will mix with the particles in the medium and form
dressed  states by shifting the mass 
and residues of the propagators. We will describe this scenario by a
sudden coupling of the bare particle(s) to the bath at an initial time
which is taken to be $t=0$.  The initial density matrix is then
factorized into a tensor 
product of a density matrix for the bare particles and another for the bath.

The  model that we choose to describe this situation  is that of an
oscillator of bare frequency $\omega_0$ (representing the physical
mass of the {\em in} particle states before the interaction) coupled
linearly with  a  bath with an infinite number of degrees of freedom
given by harmonic oscillators with frequencies $\omega_k$. Although
this 
is a drastic simplification of the theory that we are ultimately
interested in describing, this model has already been used in studies
of  dissipation\cite{ullersma}-\cite{beilok} 
decoherence\cite{unruh}, and also as models for entropy production in
heavy ion collisions\cite{elze2} and more 
recently within the context of baryogenesis\cite{yoshimura}.

The Lagrangian is given by
$$
L[q,Q_k]\,=\,\frac{1}{2}\left(\dot{q}^2\,-\,\omega_0^2q^2\right)\,+
\,\frac{1}{2}\sum_k\left( \dot{Q}_k^2\,-\,\omega_k^2~ Q_k^2
\right)\,-\,q\sum_k C_k ~Q_k, 
$$

\noindent where the different coefficients of $\dot{q}^2 ; ~
\dot{Q}^2_k$ (oscillator masses) had been absorbed by a canonical
transformation into  a redefinition of 
the couplings $C_k$. We now refer to the oscillator $q$ as the
`system' i.e. the degree of freedom whose dynamics we 
are interested in studying, and the oscillators $Q_k$ as the `bath',
these will be integrated out in the  
non-equilibrium effective action.
The degree of freedom $q$ could be associated with a particle of a
particular wave number, for example in the linear 
relaxation approximation to kinetics in which only one particular mode
is displaced from equilibrium, whereas all of the 
other modes in the field theory remain in equilibrium. In this case
the mode in question could be identified with the 
variable $q$ and the other modes in equilibrium with the bath.

 We will eventually take the limit in which the bath oscillators are
distributed continuously by introducing the bath spectral density
$J(\omega)$ and where appropriate replacing the discrete distribution 
with a continuum one in the following manner:

$$
J(\omega) = \frac{\pi}{2}\sum_k \frac{C^2_k}{\omega_k}~
\delta(\omega-\omega_k) 
$$
in such a way that 
\begin{equation}
\sum_k C^2_k ~f(\omega_k) \rightarrow \frac{2}{\pi} \; \int d\omega \;
\omega \; J(\omega)~ f(\omega). \label{specrep} 
\end{equation}

Our main goal is to study the evolution of the number of excitations
or `particle distribution' associated with the 
quanta of the system. Anticipating self-energy renormalization effects
by the medium, we define a reference frequency  
$\Omega$ and introduce the
operator that counts the number of quanta  of the system's degrees
of freedom associated with this frequency 

\begin{equation}
\hat{n}(t)\,=\,\frac{1}{2\Omega}\left[p^2(t)\,+ \,{\Omega}^2 q^2(t) 
\right]\,-\,\frac{1}{2}\; ,
\label{number}
\end{equation}
where $p(t)$ is the momentum of the particle. The reference frequency
could either be taken 
to be the bare frequency $\omega_0$, or the `in-medium' frequency
renormalized by the interaction with the bath and we will leave 
this choice unspecified for the moment.

Since the theory is quadratic we can resort to a number of different
ways to study the dynamical evolution: 
\begin{enumerate}

\item  Given an initial density matrix we can evolve it in time
exactly and obtain all of the non-equilibrium correlation functions. 

\item The Heisenberg equations of motion for the operators can be
solved exactly and again we can obtain any  correlation 
function. 

\item The normal modes can be found exactly, from which we can find
the {\em exact} ground state and also obtain the 
operators that create the particle or quasiparticle states to study the
asymptotic evolution of non-equilibrium states. 

\item We compute exactly the expectation value of the proposed number operator  in the 
canonical ensemble of the system plus bath and compare the result to the asymptotic 
form of the non-equilibrium
distribution function. This allows an unequivocal description of thermalization in terms of the density matrix.

\end{enumerate}

We will pursue all of the above different approaches, since each
particular method provides different insights and the main 
goal is to understand this simpler model in detail to provide
intuition into the more complicated  case of field theories.

\subsection{\bf Time evolution of an initial Density Matrix:}
 
 The first method is to 
calculate the time evolution of the reduced density matrix,
$\rho_r(t)$, of the particle that has been prepared at 
some initial time $t_i$.

This can be achieved by treating the infinite set of harmonic
oscillators, $Q_k$, as a `bath' 
and obtaining an influence
functional\cite{feyvern}-\cite{weiss,beilok} by tracing out the  
bath degrees of freedom. We assume that the total density matrix for the
particle-bath system decouples at the initial time $t_i$, i.e.
$$
\rho(t_i)\,=\,\rho_s(t_i)\,\otimes\,\rho_R(t_i),
$$
where $\rho_R(t_i)$ is the density matrix 
of the bath which describes infinite set of harmonic oscillators in thermal 
equilibrium at a temperature $T$ and $\rho_s(t_i)$ is the density
matrix of the particle  
which is taken to be 
that of a harmonic oscillator in thermal equilibrium at temperature
$T_0$.  More complicated initial density matrices, 
including correlations between system and bath degrees of freedom can
be studied by following the methods found in\cite{grabert}.

The complete information of non-equilibrium expectation values and correlation functions is completely 
contained in the time dependent density matrix
$$
\rho(t)= U(t,t_i)~\rho(t_i)~U^{-1}(t,t_i)
$$
with $U(t,t_i)$ the time evolution operator. Real time non-equilibrium
expectation 
values and correlation functions can be obtained via functional derivatives
with respect to sources of the generating
functional\cite{schwinger}
$$
Z[j^+,j^-] = Tr\left[ U(\infty,t_i;j^+)~\rho(t_i) ~
U^{-1}(\infty,t_i;j^-)\right]/Tr\rho(t_i), 
$$
where $j^{\pm}$ are sources coupled to the particle coordinate. This
generating functional   
is readily obtained using the Schwinger-Keldysh method which involves a path 
integral in a complex contour in time\cite{schwinger}

In the present situation, the non-equilibrium generating functional is given by
\begin{eqnarray}
&& Z[j^+,j^-]  =  \frac{1}{Tr\rho(t_i)}\int dq dq_1 dq_2
\left<q_1|\rho_s(t_i)|q_2\right> 
\int dQ dQ_1 dQ_2 \left<Q_1|\rho_R(t_i)|Q_2\right> \times \nonumber  \\
& & \int {\cal D}q^+  {\cal D}q^- {\cal D} Q^+ 
{\cal D}Q^- ~  \mbox{exp}\left\{i\int dt \left( L[q^+,Q_k^+]\,-\,
L[q^-,Q_k^-]\,-\,q^+j^+ \,+\,q^-j^-\right)\right\}\nonumber 
\end{eqnarray}
with $$dQ\,=\,\prod_k\,dQ_k\quad\quad\mbox{and}\quad\quad {\cal D}Q\,=\,\prod_k\,{\cal D}Q_k $$
and with the following boundary conditions on the fields: 
$Q^+_k(t_i)= Q_{1k} ; ~ Q^+_k(\infty)=Q^-_k(\infty)= Q_k ; ~ 
Q^-_k(t_i)=Q_{2k};~q^+(t_i)=q_1;~ q^+(\infty)=q^-(\infty)=q;~ 
q^-(t_i)=q_2$.
  The signs $\pm$ in 
the above expressions correspond to the fields and sources on
the forward ($+$) and backward $(-)$ branches. The contribution from the branch
along the imaginary time is canceled by the normalization factor. Real time, non-equilibrium Green's 
functions are now obtained as functional derivatives
with respect to the sources. There are four types of free 
propagators\cite{schwinger}
\begin{equation}
\begin{array}{lclcl}
\left< Q^+_k(t)Q^+_k(t^\prime) \right> & =& -i\,G^{++}_k(t,t^\prime) & = &   
-i\left[G^>_k(t,t')\theta(t-t')+G^<_k(t,t')\theta(t'-t)\right] \\
& & & & \vspace{-2ex}\\
\left< Q^-_k(t)Q^-_k(t^\prime) \right> & =& -i\,G^{--}_k(t,t^\prime) & = &  
-i\left[G^>_k(t,t')\theta(t'-t) + G^<_k(t,t')\theta(t-t')\right]  \\
& & & & \vspace{-2ex}\\
\left< Q^+_k(t)Q^-_k(t^\prime) \right> & =& i\,G^{+-}_k(t,t^\prime) & = &  
-i\,G^<_k(t,t')  \\
& & & & \vspace{-2ex}\\
\left< Q^-_k(t)Q^+_k(t^\prime) \right> & =& i\,G^{-+}_k(t,t^\prime) & = & 
-i\,G^>_k(t,t')= -i\,G^<_k(t',t),  
\label{greens}
\end{array}
\end{equation}
where
\begin{eqnarray}
G^>_k(t,t') & = & \frac{i}{2\omega_k} \bigg[ \left(1\,+\,N_k\right)\mbox{exp}
\left\{-i\omega_k(t-t')\right\}
\,+\,N_k\,\mbox{exp}\left\{i\omega_k(t-t')\right\} \bigg] \nonumber \\
G^<_k(t,t') & = & \frac{i}{2\omega_k} \bigg[ \left(1\,+\,N_k\right)\mbox{exp}
\left\{i\omega_k(t-t')\right\}
\,+\,N_k\,\mbox{exp}\left\{-i\omega_k(t-t')\right\} \bigg]\nonumber \\
N_k & =  & \frac{1}{ \mbox{exp}\left\{ \beta\omega_k \right\}\,-\,1 }. 
\label{Nk}
\end{eqnarray}
%
%
%
%
%
%

\subsection{The reduced density matrix}
\label{reduced}

The reduced density matrix, $\rho_r(t)$, is defined as 
$$
\rho_r(t)\,= \, \frac{{Tr}_R \rho(t)}{{Tr} \rho_R(t_i) },
$$
where the subscript $R$ in ${Tr}_R$ refers to tracing over the bath degrees of freedom. Taking the trace
over $Q_k$, one obtains the reduced density matrix in terms of the influence functional, 
${\cal F}[q^+,q^-]$
$$
\rho_r[q,q';t] \,=\, \int dq_1 dq_2~ \rho_0[q_1,q_2]
\int {\cal D}q^+  {\cal D}q^-  \mbox{exp}\left\{i\int dt \left( L_0[q^+]\,-\,
L_0[q^-]\right)\right\}\times {\cal F}[q^+,q^-],  
$$
where $\rho_0[q_1,q_2]$ is the initial density matrix of the particle and
\begin{eqnarray}
{\cal F}[q^+,q^-] & = & \mbox{exp}\left\{\frac{i}{2} \sum_k C_k^2 \int
dt \int dt' 
\sum_{a,b} q^a(t) G_k^{ab}(t,t') q^b(t') \right\}\; ;\quad a,b=+,-\nonumber \\
L_0[q^\pm] &=&
\frac{1}{2}\left[(\dot{q}^{\pm})^2\,-\,\omega_0^2(q^{\pm})^2\right]\;
. \nonumber 
\end{eqnarray}

We will choose the initial density matrix of the particle to be that of 
an harmonic oscillator of reference frequency $\Omega$ in thermal
equilibrium at temperature $T_0$  given  
by
\begin{eqnarray}
\rho_0[q_1,q_2]&=&\sqrt{\frac{1}{2\pi\sigma}}   \, \mbox{exp} 
\left\{ ip_i (q_1-q_2)\right\} \nonumber\\
& & \times \, \mbox{exp}\left\{-\frac{\Omega}{2
\mbox{sinh}\left[\beta_0\Omega\right]}\Big[\left(
(q_1-q_i)^2+(q_2-q_i)^2 \right) 
\mbox{cosh}\left[\beta_0 \Omega \right]\,-\,2(q_1-q_i)(q_2-q_i) \Big]
\right\},\nonumber 
\end{eqnarray}
where $q_i$ and $p_i$ are respectively the average position and
momentum of the particle, 
$\beta_0=1/T_0$ and
$$
\sigma \,=\, \frac{1}{2  \Omega} \mbox{coth}\left[\frac{\beta_0
\Omega}{2}\right] = \frac{1+2n(0)}{2  \Omega}\; \; ; \; \;  
n(0) = \frac{1}{e^{\beta_0 \Omega}-1}. 
$$

The reference frequency $\Omega$ will allow to understand the different
features of the dynamics of the
dressed particle in the medium, rather than the bare particle with
frequency $\omega_0$. We will specify this reference frequency below when we study the dynamics in detail.

At this stage it is convenient to introduce the center of mass and relative 
coordinates, $x$ and $r$ respectively, which are defined as
$$
x(t')\,=\, \frac{1}{2} \left[ q^+(t')\,+\,q^-(t') \right]
\hspace{.3in},\hspace{.3in} 
r(t')\,=\,q^+(t')\,-\,q^-(t').
$$
These are recognized as the Wigner
coordinates\cite{caldeira}-\cite{weiss}. In terms of  
these coordinates the reduced density matrix becomes
\begin{equation}
\rho_r[x_f,r_f;t] \,=\, \int dx_i \, dr_i \, \rho_0[x_i,r_i] \, e^{i
\left(\dot{x}_f r_f 
- \dot{x}_i r_i \right) } \int {\cal D}x  {\cal D}r \, e^{i S_{eff} }, 
\label{reddenmat}
\end{equation}
where
\begin{eqnarray}
x_i & = & \frac{1}{2} \left( q_1\,+\,q_2 \right) \hspace{1em};\hspace{1em}
r_i\,=\,q_1\,-\,q_2 \hspace{1em};\hspace{1em}
x_f  =  \frac{1}{2} \left( q\,+\,q' \right) \hspace{1em};\hspace{1em}
r_f\,=\,q\,-\,q' \nonumber\\
\rho_0[x_i,r_i]&=&\sqrt{\frac{1}{2\pi\sigma}}\exp\left\{-\frac{1}{2\sigma}
\left( x_i^2 - 2 q_ix_i+q_i^2 \right) - \frac{\Omega^2\sigma}{2} r_i^2
+ ip_ir_i \right\} \nonumber\\
S_{eff}&=&\int_{0}^t dt' \left\{- \left[ \ddot{x}(t') + \omega_0^2~
x(t') \right]r(t') 
\right.\nonumber\\
& & +\,\left.\int_{0}^t dt'' \left[ r(t')~ \Sigma(t'-t'')~
x(t'')\,+\,\frac{i}{2} r(t')~ 
K(t'-t'')~ r(t'') \label{seff}
\right] \right\}
\end{eqnarray}
with
\begin{eqnarray}
\Sigma(t'-t'') & = & \theta(t'-t'') \sum_k \frac{C_k^2}{\omega_k} \sin\left[
\omega_k (t'-t'') \right] \label{bigsigma} \\
K(t'-t'') & = & \sum_k \frac{C_k^2}{2 \omega_k} 
\mbox{coth}\left[\frac{\beta\omega_k}{2} \right] \cos\left[\omega_k
(t'-t'')\right]\; .\label{Kkernel} \\ \nonumber
\end{eqnarray}
The kernel $\Sigma(t'-t'')$ is  the retarded self energy of the system
degree of freedom $ q $. 

Since $S_{eff}$ is gaussian, the path integral can be evaluated
exactly using the saddle  
point method where the coordinates $x(t')$ and $r(t')$ are expressed
as variations  
around the extremum configurations $\bar{x}(t')$ and $\bar{r}(t')$
respectively.  
These satisfy the following equations of motion\cite{grabert}

\begin{equation}
\ddot{\bar{x}}(t')+\omega_{0}^{2}~\bar{x}(t')\:-\:
\sum_{k}\frac{C_{k}^{2}}{\omega_{k}}\int_{0}^{t'}dt''
\sin[\omega_{k}(t'-t'')]\:\bar{x}(t'')\:=\:0
\label{xbar} 
\end{equation}
\begin{equation}
\ddot{\bar{r}}(t')+\omega_{0}^{2}~\bar{r}(t')\:+
\:\sum_{k}\frac{C_{k}^{2}}{\omega_{k}}
\int_{t'}^{t}dt''\sin[\omega_{k}(t'-t'')]\:\bar{r}(t'')\:=\:0
\label{rbar} 
\end{equation}
with
\[ \bar{x}(0)\;=\;x_{i} \quad ; \quad  \bar{x}(t)\;=\;x_{f} \quad ; \quad
 \bar{r}(0)\;=\;r_{i} \quad ; \quad \bar{r}(t)\;=\;r_{f}. \]

 There is in principle an imaginary inhomogeneity in the equation of
motion for the center of mass variable (\ref{xbar}), 
but this inhomogeneity is shown to be a surface term and combines with
the end point contributions. Therefore only the 
real part of the classical trajectory is relevant.(For details see ref.
\cite{grabert} page 145.) 
The eqs.(\ref{xbar}) and (\ref{rbar}) can be solved by Laplace
transform, and we obtain  
\begin{eqnarray}
\bar{x}(t')& = &  x_i \dot{g}(t')\,+\, \dot{x}_i g(t') \nonumber \\
\bar{r}(t')& = &  r_i \, \frac{g(t-t')}{g(t)} \,+\, r_f \,
\frac{{g}^-(t')}{g^-(t)}\nonumber \\ 
g^-(t') &\equiv& \frac{\dot{g}(t)\,g(t-t')\,-\,g(t)\,\dot{g}(t-t')}{
g(t)\,\ddot{g}(t)\,-\,\dot{g}^2(t)},  \nonumber
\end{eqnarray}
where the  function $g(t)$ and its first derivative $\dot{g}(t)$ are
solutions of (\ref{xbar}) with initial 
conditions $g(0)=\ddot{g}(0)=0$ and $\dot{g}(0)=1$ and $\dot{x}_i$ 
is obtained from $x_i\; , x_f$ and $g(t)$. Such solution 
 is given by the inverse Laplace transform of 
$\tilde{g}(s)$ which is given by
\begin{equation}
\tilde{g}(s) \,=\,\frac{1}{s^2\,+\,\omega_0^2\,+\,\tilde{\Sigma}(s) }
\label{glap}
\end{equation}
with the Laplace transform of the retarded self energy given by
\begin{equation}
\tilde{\Sigma}(s) \,=\, -\sum_k
\frac{C_k^2}{\omega_k}\frac{\omega_k}{s^2+\omega_k^2} \rightarrow 
-\frac{2}{\pi} \int d\omega J(\omega) \frac{\omega}{s^2+\omega^2}\; .
\label{selfenergy} 
\end{equation}
Here we have taken the limit of a continuum distribution of bath
oscillators as given by eq.(\ref{specrep}). 

We will postpone the computation of the function $g(t)$ to the next
section and  we will provide a particular spectral density for the
bath in a later section wherein we will compare exact results to different
approximations.  

Finally, substituting $\bar{x}(t')$ and $\bar{r}(t')$  and 
evaluating the remaining gaussian integrals in eq.(\ref{reddenmat}) we are
led to the reduced density matrix 
\begin{eqnarray}
\rho_r[x_f,r_f;t]& = &{\frac{1}{2\,{\sqrt{\pi\,A(t)}}}} \, \mbox{exp}
\left\{-\frac{ 
1}{2}\left[{\frac{\,\sigma }{\left(g^-(t)\right)^2}} + {R^{--}(t)}  
   -\,{\frac{ B^2(t) }{2\,A(t)}}\right] r_f^2\,-\,\frac{1}{4\,A(t)}x_f^2
\right.\nonumber \\
& & +\,\,i  \left[ \frac{\dot{g}(t)}{g(t)}-\frac{B(t)}{2\,A(t)}
\right]\,x_f\,r_f\, 
+\, i  \left[\frac{B(t)}{2\,A(t)}\Big(p_i\,g(t)+ q_i
\dot{g}(t)\Big)\,-\,\frac{q_i}{g^-(t)} 
\right]\,r_f \nonumber \\
& & \left. +\,\,\frac{1}{2\,A(t)} \Big(p_i g(t)+ q_i \dot{g}(t)\Big)\,x_f\,-\,
\frac{1}{4\,A(t)} \Big(p_i g(t)+ q_i \dot{g}(t)\Big)^2 \right\}\;,\nonumber
\end{eqnarray}
where
\begin{eqnarray}
A(t)&\equiv& \frac{\Omega^2\,\sigma}{2}g^2(t)\,+\,\frac{1}{2}R^{++}(t)\,+\,
\frac{\sigma}{2}\dot{g}^2(t)\nonumber \\
B(t) & \equiv & \frac{\sigma }{g^-(t)}\dot{g}(t)\,-\,R^{+-}(t) \nonumber \\
R^{++}(t) &\equiv& \int_0^t dt' \int_0^t dt''\,g(t-t') \,
K(t'-t'')\, g(t-t'') \nonumber \\
R^{--}(t) &\equiv& \int_0^t dt' \int_0^t dt''\,\frac{g^-(t')}{g^-(t)} \,
K(t'-t'')\, \frac{g^-(t'')}{g^-(t)} \nonumber \\
R^{+-}(t) &\equiv& \int_0^t dt' \int_0^t dt''\,g(t-t')\, K(t'-t'') \,
\frac{g^-(t'')}{g^-(t)}\; .  \nonumber
\end{eqnarray}

Having obtained the reduced density matrix, we can now obtain the expectation values of $q^2(t)$ and 
$p^2(t)$ and to compute the expectation value of the number operator
(\ref{number}), which after some straightforward algebra is shown to
be  given by 
\begin{eqnarray}
\left<n(t)\right> & = & -\frac{1}{2} \,+\, {\cal R}(t)\,+\,\frac{\Omega}{2}
\left(p_i^2+ \Omega^2 \sigma \right)\,g^2(t)
\,+\,\frac{q_i^2+ \sigma }{2\,\Omega}\,\ddot{g}^2(t) \nonumber \\
& & +\,\,\frac{p_i^2\,+ \Omega^2 \left(q_i^2\,+ 2 \sigma
\right)}{2\,\Omega}\,\dot{g}^2(t)\,+\, 
\frac{p_i q_i}{\Omega} \left[\ddot{g}(t)+\Omega^2 g(t) \right]
\,\dot{g}(t)\; ,
\label{exactn}
\end{eqnarray}
where we have introduced the shorthand notation 
$$
{\cal R}(t)\,\equiv\, \frac{1}{2\,\Omega} \left\{
R^{--}(t)\,+\, 2 \, \frac{\dot{g}(t)}{g(t)}\,R^{+-}(t)\,+\,\left[
\frac{\dot{g}^2(t)}{g^2(t)}\,+\,\Omega^2 \right]\,
R^{++}(t) \right\}\; .
$$
The expression for ${\cal R}(t)$ can be simplified by introducing the functions
\begin{eqnarray}\label{anterior}
h(\omega,t) & \equiv & \int_0^t d\tau \, e^{-i \omega \tau} \, g(\tau)
\nonumber \\ 
k(\omega,t) & \equiv & \int_0^t d\tau \, e^{-i \omega \tau} \, \dot{g}(\tau) \\
 & = & i \omega \, h(\omega,t) + e^{-i \omega t} g(t). \nonumber 
\label{hkwt}
\end{eqnarray}
In terms of these functions, ${\cal R}(t)$ can be written as
\begin{equation}\label{siguiente}
{\cal R}(t) \,=\,\frac{1}{4  \Omega} \sum_k \frac{C_k^2}{\omega_k}
\left[1+2N(\omega_k)\right]\left(\,|k(\omega_k,t)|^{2} + \Omega^2 \,
|h(\omega_k,t)|^{2}\right) 
\end{equation}
and in the limit of a continuum spectrum of bath oscillators
$$
{\cal R}(t) \,=\,\frac{1}{2\pi  \Omega } \int d\omega J(\omega)
\left[1+2N(\omega)\right] \left(\,|k(\omega,t)|^{2} + \Omega^2 \,
|h(\omega,t)|^{2}\right). 
$$

The expectation value of the number operator (\ref{exactn}) in the 
non-equilibrium density matrix has two contributions: one that is completely
determined by the initial state of the system (proportional to $p_i\; ; \; q_i \; ; \; \sigma$) and the other, determined by the bath and given by 
${\cal R}(t)$. 

\subsection{Fluctuation-Dissipation}
The main advantage of studying the time evolution of the reduced density
matrix is that it allows to establish a direct relationship between
the relaxation of the occupation number of the system and the
fluctuation dissipation theorem. The connection between the fluctuation-dissipation and the Boltzmann equation
has been investigated recently in the semi-classical regime\cite{greiner}. 

This relationship is established by  re-writing the term quadratic in the
relative variable $r(t)$ in $S_{eff}$ given by the last term in
eq.(\ref{seff}) in the following form 

\begin{eqnarray}
\exp\left\{-\frac{1}{2} \int dt \int dt'\; r(t)\; K(t-t')\; r(t')\right\} & = &
 \int {\cal D}\xi \exp\left\{-\frac{1}{2} \int dt \int dt'\; \xi(t)\;
 K^{-1}(t-t')\; \xi(t')  \right. \nonumber \\ 
&+ & \left. i \int dt\; \xi(t)\; r(t) \right\}\; . \nonumber
\end{eqnarray}

The path integrals in eq.(\ref{reddenmat}) can now be written in the
following form\cite{schmid}
\begin{eqnarray}
\int {\cal D}x {\cal D}r\; e^{iS_{eff}(x,r)}& = &  \int {\cal D}x
{\cal D}r {\cal D} \xi\; P[\xi]\; e^{i\tilde{S}_{eff}(x,r,\xi)} \nonumber
\\ 
\tilde{S}_{eff}(x,r,\xi) & = & \int^t_{t_0}dt'\;r(t')\left\{
-\ddot{x}(t')-\omega^2_0\; x(t') + \int dt''  \;
\Sigma(t'-t'')\;x(t'')+\xi(t') \right\} \; , \nonumber \\
P[\xi] & = & \exp\left\{-\frac{1}{2} \int dt \int dt'\; \xi(t)\;
K^{-1}(t-t')\; \xi(t')\right\}\; . \nonumber
\end{eqnarray}
The path integral over the relative coordinate leads to a non-Markovian
Langevin equation for $x(t)$ in the presence of a stochastic Gaussian
(but colored) noise term $\xi$. The noise correlation function is determined by $K(t-t')$ given by eq.(\ref{Kkernel}). 

The fluctuation-dissipation relation is established in the following manner.
In the limit of a
continuum distribution of the bath oscillators we find the time Fourier transform of the retarded
self-energy $\Sigma(t)$ (\ref{bigsigma}) to be given by the analytic 
continuation of the Laplace transform (\ref{selfenergy}) $s \rightarrow
\omega-i\epsilon $, i.e. 
$$
\tilde{\Sigma}(\omega-i\epsilon) = -\frac{2}{\pi} \int d\omega' 
\frac{\omega' ~ J(\omega')}{\left[
(\omega')^2-(\omega-i\epsilon)^2\right]}\; . 
$$
 Then we
find (for $\omega >0$) 
$$
\mbox{Im}\left[\tilde{\Sigma}(\omega)\right] = J(\omega) 
$$
and the Fourier transform in time of the kernel $K(t)$ (\ref{Kkernel})
is given by 
$$
\tilde{K}(\omega) = \frac{1}{2\pi}
\mbox{Im}\left[\tilde{\Sigma}(\omega)\right] \coth\left[\frac{\beta
\omega}{2}\right]\; .
$$
This is the usual fluctuation-dissipation relation\cite{weiss}.
Finally we obtain the bath contribution to the non-equilibrium
occupation number, which is determined by ${\cal R}(t)$ in a form
that displays clearly its relationship to the fluctuation dissipation
relation
$$
{\cal R}(t) \,=\,\frac{1}{  \Omega } \int d\omega \tilde{K}(\omega)
\left(\,|k(\omega,t)|^{2} + \Omega^2 \, |h(\omega,t)|^{2}\right),
$$
\noindent where $\tilde{K}(\omega)$ is the power spectrum of the
 bath. This expression makes explicit the stochastic nature of 
 thermalization and establishes a direct relationship with the fluctuation dissipation theorem.

Detailed understanding of the particle number relaxation  requires the knowledge of the dynamical function $g(t)$
 which  will be studied in a later section for a particular choice of the spectral density of the bath. 

%
%
%
%
%
%

\subsection{The Heisenberg Operators}
\label{operator}
The above results can be understood in an alternative manner by obtaining the real
time evolution of the Heisenberg picture operators, from which the expectation value of the number operator can be obtained by providing an
initial density matrix. 
 
The equations of motion of the Heisenberg operators $q(t)$, $p(t)$ and
$Q_k(t)$  are given by 
\begin{eqnarray}
\ddot{q}(t) + \omega_{0}^{2}\;q(t) & =& -\sum_{k}{C_{k}}\;Q_{k}(t)
\label{eqna} \\ 
\ddot{Q_{k}}(t)+\omega_{k}^{2}\;Q_{k}(t) & =& -\; q(t)\; {C_{k}}. \label{eqnb}
\end{eqnarray} 
Solving eq.(\ref{eqnb}) using the retarded Greens function of the operator $d^2/dt^2 +
\omega^2$ given by
$$ 
G(t,t')\,=\,\frac{1}{\omega} \sin\left[\omega (t-t')\right]\;
\theta(t-t'), 
$$ 
we find
$$
Q_{k}(t)\;=\;Q_{k}^{(0)}(t)\; -\;
\frac{C_{k}}{\omega_{k}}\int_{0}^{t}dt'\;
\sin\left[\omega_{k}(t-t')\right]\;q(t')\;,
$$
where $Q_{k}^{(0)}(t)$ is the homogeneous solution of
eq.(\ref{eqnb}). Substituting  
$Q_{k}(t)$ back in eq.(\ref{eqna}), we obtain the equation of motion
of the operator $q(t)$  
$$
\ddot{q}(t) + \omega_{0}^{2}\; q(t)
-\sum_{k}\frac{C_{k}^{2}}{\omega_{k}}\int_{0}^{t}dt'\;
\sin\left[\omega_{k}(t-t')\right]
\;q(t')\;=\;-\;\sum_{k}{C_{k}}\;Q_{k}^{(0)}(t)\; .
$$

\noindent Taking the expectation value of the above equation in the
initial density matrix we recognize the equation of motion for the 
center of mass Wigner variable, eq.(\ref{xbar}) obtained in the
evolution of the density matrix 

The above equation is solved using Laplace transform, and the operator
solution with the initial condition 
$q(t=0)=q(0)\; ; \; \dot{q}(t=0)=p(0)$ is found to be given by
\begin{equation}
q(t)\;=\;p(0)\;g(t)+q(0)\;\dot{g}(t)-\sum_{k}{C_{k}}\int_{0}^{t}d\tau
\; Q_{k}^{(0)}(t-\tau)\;g(\tau)\;, 
\label{qinlap}
\end{equation}
where $g(t)$ is the same  function which was defined in the previous
section and whose Laplace transform is given by eq.(\ref{glap}).

It is convenient to write
$$
Q_{k}^{(0)}(t)\;=\;\frac{1}{\sqrt{2\omega_{k}}}\left[a_{k}\,e^{-i\omega_{k}t}\,
+ \, a_{k}^{\dag}\,e^{i\omega_{k}t}\right]\; ,   
$$
so that the integral in the last term in eq.(\ref{qinlap}) becomes
$$
\int_{0}^{t} d\tau\;
Q^{(0)}_{k}(t-\tau)\;g(\tau)\;=\;\frac{1}{\sqrt{2\omega_{k}}} 
\left[a_{k}^{\dag}\;e^{i\omega_{k}t}h(\omega_{k},t)+h.c.\right]\; , 
$$
where $h(\omega,t)$ is defined in eq.(\ref{anterior}). 

Since the initial density matrix describes a thermal distribution for
the quanta of a harmonic oscillator of reference frequency $\Omega$, it
is convenient to write the initial position and momentum operators  in terms of the creation and
annihilation operators of a quanta of frequency $\Omega$ as
\[q(0)\;=\;\frac{1}{\sqrt{2\Omega}}\left[b \;+\;
b^{\dag}\right]\quad;\quad p(0)\;=\; 
-i\;\sqrt{\frac{\Omega}{2}}\left[b\;-\;b^{\dag}\right]. \]
Gathering all terms, $q(t)$ and $p(t)$ become
\begin{eqnarray}
q(t) &=& \frac{1}{\sqrt{2\Omega}}\left[ b\left(\dot{g}(t)-i\Omega
g(t)\right) + b^{\dag}\left(\dot{g}(t)+i\Omega
g(t)\right)\right]-\sum_{k}\frac{C_{k}}{\sqrt{\omega_{k}}}
\left[a_{k}^{\dag}\,e^{i\omega_{k}t}\,
h(\omega_{k},t)+h.c.\right], \nonumber \\
p(t) &=&
\frac{1}{\sqrt{2\Omega}}\left[b\left(\ddot{g}(t)-i\Omega\dot{g}(t)\right)
+ b^{\dag}\left(\ddot{g}(t)+i\Omega \dot{g}(t) 
\right)\right] - \sum_{k}\frac{C_{k}}{\sqrt{\omega_{k}}}\left[a_{k}^{\dag}\;
e^{i\omega_{k}t}k(\omega_{k},t)+h.c.\right]\; ,\nonumber \\ \nonumber
\end{eqnarray}
where $h(\omega_k,t) ~; k(\omega_k,t)$ are defined in eq.(\ref{anterior}).
The expectation value of the occupation number operator ${\hat n}(t)$ in
eq.(\ref{number}) can be  
evaluated using an initial density matrix which is diagonal in the basis
of the bare number operators for system and bath. Assuming a continuum
spectrum of the bath  oscillators, using eq.(\ref{specrep}) and
considering 
for simplicity the case of vanishing expectation values of $q(0)\; ; \; p(0)$
in the initial density matrix, we find 
\begin{eqnarray}
\left<n(t)\right> & =& -\,\frac{1}{2} \,+\,
\frac{1+2n(0)}{4\Omega^{2}} \left[\ddot{g}^{2}(t)+2
\Omega^{2}\dot{g}^{2}(t)+ \Omega^{4}g^{2}(t) 
\right] \nonumber \\
& & + \quad  \frac{1}{2\pi \Omega}
\int d\omega \; {J(\omega)}
\left[1+2N(\omega)\right]\left[\,|k(\omega,t)|^{2}+ \Omega^2 \,
|h(\omega,t)|^{2}\right] \; . \label{opnumb}  
\end{eqnarray}
Setting $q_i=0 \; ; \; p_i=0$ in the result (\ref{exactn}) 
we find that eq.(\ref{opnumb}) reduces to the expression obtained by the
time evolution of the density matrix (\ref{exactn}) and the last term
is identified with ${\cal R}(t)$.

The operator method allows to compute any correlation function of
operators in the initial density matrix at arbitrary times, whereas the
time evolution of the density matrix would require the introduction of
external sources and taking functional derivatives with respect to those to
obtain unequal time correlation functions.
 
%
%
%
%
%
%

\section{Real Time Evolution: $g(t)$ and $n(t)$ \label{secgt}}

Before specifying a choice of the spectral density of the bath
$J(\omega)$ we can obtain more insight by analyzing the real time 
behavior of $g(t)$ and consequently of $<n(t)>$ in general. Having
determined the general features 
of the evolution, we will then specify a particular choice of
$J(\omega)$ and provide a detailed numerical study comparing with
different approximations in a later section. In general the spectral
density fullfils 
$$
J(\omega)= \left\{ \begin{array}{cc}
\neq 0 & \mbox{for \,  $\omega_{th} < |\omega| < \omega_c$} \\
0 & \mbox{otherwise}
\end{array} 
\right.
$$
where $\omega_{th} \; ; \omega_c$ are threshold and cutoff frequencies 
respectively. 

The real time evolution of $g(t)$ is given by the inverse Laplace transform 

\begin{equation}
g(t) = \frac{1}{2\pi i} \int_\Gamma e^{st} \; \tilde{g}(s)  \;ds \;,
\label{contint}
\end{equation}
where $\tilde{g}(s)$ is given by eq.(\ref{glap}) and $\Gamma$ refers to the Bromwich 
contour running along the imaginary
axis to the right of all the singularities of $\tilde{g}(s)$ in the complex $s$ plane. Therefore we need to understand the analytic structure of $\tilde{g}(s)$ to obtain the 
real time dynamics of the particle occupation number. 

>From the  expression (\ref{selfenergy}) for the Laplace transform of the
retarded self-energy, we find that 
$\tilde{\Sigma}_S(s)$ has cuts along the imaginary $s$-axis 
for  $s=i\omega \; ; \omega_{th} < |\omega| < \omega_c$ as can be
seen from 
$$
\tilde{\Sigma}_S(s=i\omega\pm 0^+)\,=\,\Sigma_R(\omega)\,\pm \,i\,
\Sigma_I(\omega) 
$$
with
\begin{eqnarray}
\Sigma_R(\omega) & = & \frac{2}{\pi } \, {\cal P}\int d\omega'\; 
\frac{\omega'\;J(\omega')}{{\omega}^2-{\omega'}^2} \label{realsig} \\
\Sigma_I(\omega)& = & {2}\:\mbox{sign}(\omega)\:\int d\omega'
J(\omega')\;  \omega'\;
\delta({\omega'}^{2}-\omega^{2}) \nonumber\\ 
 & = &  \, \mbox{sign}(\omega)\; {J(|\omega|)}\;.\label{imsig} 
\end{eqnarray}

 As in field theory, it is convenient to introduce a renormalized 
frequency by performing a subtraction of the self-energy. Clearly the
subtraction point is arbitrary, and we choose to subtract at $s=0$. 
We thus introduce the renormalized frequency as 
\begin{equation}
\omega^2_R\,= \,\omega^2_0 + \tilde{\Sigma}(s=0)=
\,\omega^2_0 \,-\,\frac{2}{\pi } \int d\omega\;
\frac{J(\omega)}{\omega}\label{renofreq}\;, 
\end{equation}
and the once subtracted self energy is given by
$$
\tilde{\Sigma}_S(s)\,=\, {\Sigma}(s)-{\Sigma}(s=0) = \,\frac{2}{\pi }
\int_0^\infty d\omega \;\frac{J(\omega)}{\omega} \;
\frac{s^2}{s^2\,+\,\omega^2 }\;.
$$

Isolated poles of $\tilde{g}(s)$ are at the values $s_p$ which satisfy 
$$
s_p^2\,+\,\omega_R^2\,+\,\tilde{\Sigma}_S(s_p) \,=\, 0.
$$
\noindent These are purely imaginary when they are 
below the threshold of the bath. These correspond to exact stable
states of the particle-bath interacting system.  

If the imaginary part of the pole (in the $s$-variable) $\omega_p$ is
above threshold ($\omega_p > \omega_{th}$,  then the pole is in the
second (unphysical) Riemann sheet and for 
weak couplings the spectral density $S(\omega)$ will feature a
Breit-Wigner resonance shape 
where the width of the resonance is related to the imaginary part of
the kernel $\tilde{\Sigma}_S$ and the peak of the resonance is at
$\omega_p$. 

The position of these complex poles can be parametrized in terms of
real and imaginary parts as  
$$
s_p= i\omega_p -\Gamma. \label{poles}
$$

These correspond to unstable states and are not eigenstates of the
interacting Hamiltonian. If the width $\Gamma << \omega_p$ these are 
long-lived resonances and are {\em almost} energy eigenstates. 
These states will be identified with  quasiparticles in the next section. 

Depending on the strength of the coupling with the environment,
$J(\omega)$, and the value  
of $\omega_R$, the imaginary part of the pole, $\omega_p$, can be
above or below the threshold, $\omega_{th}$.

\begin{figure}
\begin{large}
\fbox{
\setlength{\unitlength}{1mm}
\begin{picture}(150,90)
\linethickness{0.45pt}
\put(5,45){\line(1,0){60}}
\put(35,0){\line(0,1){90}}
\put(32,5){\line(0,1){25}}
\put(38,5){\line(0,1){25}}
\put(32,60){\line(0,1){25}}
\put(38,60){\line(0,1){25}}
\qbezier (32,30)(35,34)(38,30)
\qbezier (32,60)(35,56)(38,60)
\qbezier (32,5)(35,1)(38,5)
\qbezier (32,85)(35,89)(38,85)
\put(35,52){\circle{6.5}}
\put(33.35,50.5){$\times$}
\put(35,38){\circle{6.5}}
\put(33.35,36.5){$\times$}
\put(10,75){\mbox{a)}}
\put(80,75){\mbox{b)}}
\put(40,29){\mbox{$- i \omega_{th} $}}
\put(40,59){\mbox{$+ i \omega_{th} $}}
\put(40,4){\mbox{$- i \omega_{c} $}}
\put(40,84){\mbox{$+ i \omega_{c} $}}
\thicklines
\put(32,18){\vector(0,-1){0}}
\put(32,70){\vector(0,-1){0}}
\put(38,20){\vector(0,1){0}}
\put(38,72){\vector(0,1){0}}

\put(36,34.5){\vector(1,0){0}}
\put(34,41.5){\vector(-1,0){0}}
\put(36,48.5){\vector(1,0){0}}
\put(34,55.25){\vector(-1,0){0}}

\linethickness{2.0pt}
\put(35,60){\line(0,1){25}}
\put(35,5){\line(0,1){25}}
\put(34,30){\line(1,0){2}}
\put(34,60){\line(1,0){2}}
\put(34,5){\line(1,0){2}}
\put(34,85){\line(1,0){2}}

\linethickness{0.45pt}
\put(75,45){\line(1,0){60}}
\put(105,0){\line(0,1){90}}
\put(102,5){\line(0,1){25}}
\put(108,5){\line(0,1){25}}
\put(102,60){\line(0,1){25}}
\put(108,60){\line(0,1){25}}
\qbezier (102,30)(105,34)(108,30)
\qbezier (102,60)(105,56)(108,60)
\qbezier (102,5)(105,1)(108,5)
\qbezier (102,85)(105,89)(108,85)
\put(130,80){\mbox{$s$-plane}}
\put(110,29){\mbox{$- i \omega_{th} $}}
\put(110,59){\mbox{$+ i \omega_{th} $}}
\put(110,4){\mbox{$- i \omega_{c} $}}
\put(110,84){\mbox{$+ i \omega_{c} $}}
\thicklines
\put(102,18){\vector(0,-1){0}}
\put(102,70){\vector(0,-1){0}}
\put(108,20){\vector(0,1){0}}
\put(108,72){\vector(0,1){0}}

\linethickness{2.0pt}
\put(105,60){\line(0,1){25}}
\put(105,5){\line(0,1){25}}
\put(104,30){\line(1,0){2}}
\put(104,60){\line(1,0){2}}
\put(104,5){\line(1,0){2}}
\put(104,85){\line(1,0){2}}

\end{picture}}
\end{large}
\caption{ The complex contour used to evaluate $g(t)$ for the cases in which a) the pole is 
below the threshold and b) the pole is above  threshold. \label{figcont}}
\end{figure}
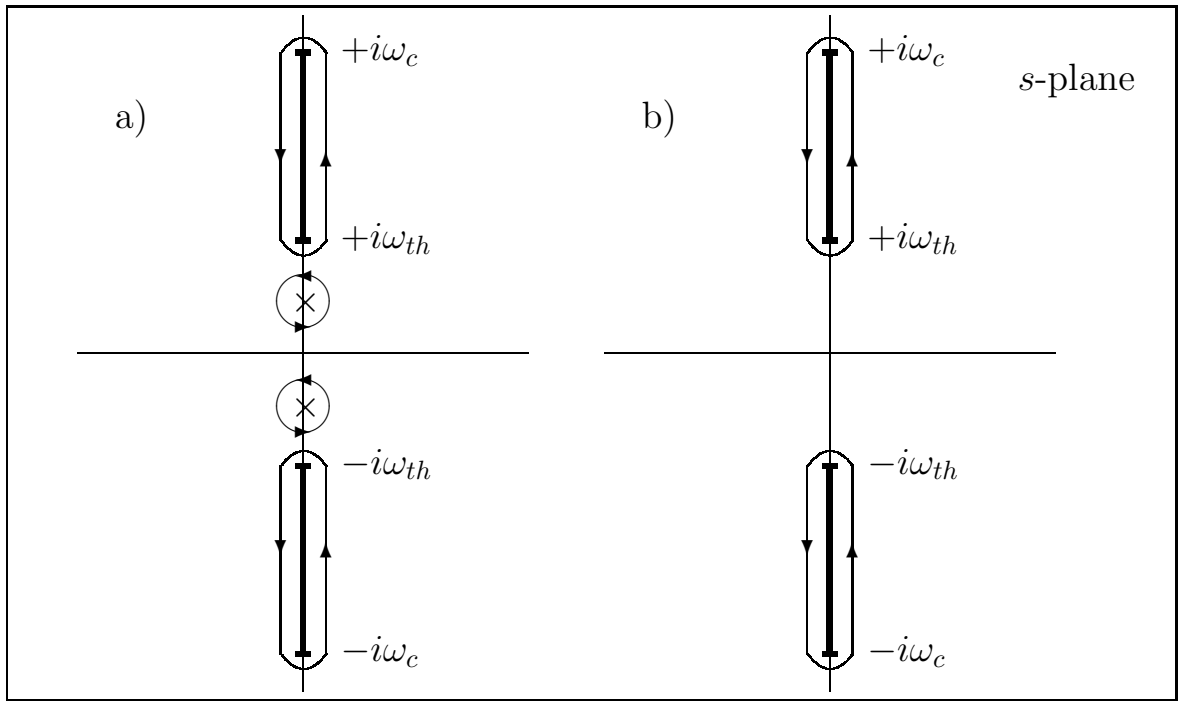

\noindent {\bf I)} Consider first the case in which the pole is above
threshold, i.e. $\omega_p > \omega_{th}$. Since there are no isolated
singularities below threshold,  only the cut will contribute to the
integral (\ref{contint}).The Bromwich contour  
$\Gamma$ in the complex $s$-plane is chosen as the one shown in
fig. \ref{figcont} where all  
the singularities of $\tilde{g}(s)$ are to the left of the
contour. Evaluating the integral  along the contour, we obtain
$$
g(t)\:=\:\frac{2}{\pi}\int_{\omega_{th}}^{\omega_{c}}d\omega\,S(\omega)
\,\sin(\omega t), 
$$
where  the spectral density  $ S(\omega) $ is given by
\begin{equation}
S(\omega)  =   \Sigma_I(\omega)\,\left| \tilde{g}(s=i\omega +
 \epsilon)\right|^2   =
 \frac{\Sigma_{I}(\omega)}{[\omega^{2}-\omega^{2}_{R}-\Sigma_{R}(\omega)]^{2}
+[\Sigma_{I}(\omega)]^{2}}\; . \label{specdef}      
\end{equation}
>From the initial condition $\dot{g}(0)=1$ we find the sum rule
\begin{equation}
\frac{2}{\pi}\int_{\omega_{th}}^{\omega_{c}}d\omega\,S(\omega) =1\;.
\label{sumruleabovecut}
\end{equation} 

For weak coupling, the spectral density can be approximated by  
a Breit-Wigner resonance and asymptotically $g(t)$ is approximately
given by\cite{attanasio} 
\begin{equation}
\dot{g}(t) \sim Z\; {\cos(\omega_p t+\alpha)}\; e^{-\Gamma \,t} \quad ;
\quad \Gamma \sim 
\frac{Z\Sigma_I(\omega_p)}{2 \omega_p} \quad ; \quad
Z= \left[1-\frac{\partial \Sigma_R(\omega)}{\partial
\omega^2}\right]^{-1}_{\omega=\omega_p} \label{abovecut} 
\end{equation}

\noindent with $\alpha$ a constant phase-shift\cite{attanasio}. We
identify this behavior with a typical quasiparticle which acquires
a width through medium effects and whose residue at the quasiparticle
pole is smaller than one as a consequence of the overlap between the
initial bare particle state and the continuum of the bath. This
interpretation will be further clarified when we study the exact
normal modes in the next section.

\noindent {\bf II)} Consider next the case in which there is only a single isolated pole below the cut. In this case, there
are two contributions to the integral (\ref{contint}); the pole
contribution and the  cut contribution. In this case we find  
\begin{equation}
\dot{g}(t)\:=\:{Z}~{\cos(\omega_{p}t)}+\frac{2}{\pi}
\int_{\omega_{th}}^{\omega_{c}}d\omega\,\omega \, S(\omega)
\,\cos(\omega t)\label{polebelow}\;, 
\end{equation}
where we define the wave function renormalization $Z$ as in
(\ref{abovecut}) above,  
\begin{equation}
Z= \left[1-\frac{\partial \Sigma_R(\omega)}{\partial
\omega^2}\right]^{-1}_{\omega=\omega_p}\label{wavefunc}. 
\end{equation}
Asymptotically at long time, the cut contribution vanishes with a
power law determined by the behavior of $S(\omega)$ near
threshold\cite{attanasio}, 
and $g(t)$ oscillates with the pole frequency $\omega_p$. Just as in
the previous 
case, the bare particle has been dressed by the medium effects, and to
distinguish from the bare or quasiparticle 
we call this state the dressed particle. The position of the dressed
particle pole has been shifted and its residue  
 is smaller than one as a result of the overlap
with the continuum of states of the bath.

>From the initial condition $ \dot{g}(0)\,=\,1 $, we derive the
important sum rule  
\begin{equation}
{Z} \, + \,
\frac{2}{\pi}\int_{\omega_{th}}^{\omega_{c}}d\omega\,\omega\,
S(\omega) 
\,=\,1 \label{sumrulebelowcut}.
\end{equation}
Both cases of the sum rule
(\ref{sumruleabovecut}) and (\ref{sumrulebelowcut}) are a consequence of the
canonical commutation relations.  
Since the spectral density $S(\omega)$ is positive semi-definite, the
above sum rule  determines that $Z \leq 1$.

The expression (\ref{polebelow}) allows us to explore the concept of
the dressing time of the particle. At long 
times the contribution  to $g(t)$ from the continuum vanishes
typically as a power law determined by the behavior of the 
spectral density near threshold\cite{attanasio} and the contribution
from the pole dominates the dynamics. This contribution results in a
asymptotic oscillatory behavior of $\dot{g}(t)$ with an amplitude
determined by the residue $Z$ at the particle pole. The formation time
can be defined to be the 
time it takes for the amplitude of $\dot{g}(t)$ to reach 
its asymptotic value $Z$ (initially $\dot{g}(0)=1$). In the case in
which the pole is embedded in the continuum (unphysical Riemann sheet) and we deal with quasiparticles,
a similar concept can be introduced, now being the formation time of the quasiparticle. There are now
two competing time scales: the formation time scale during
which the quasiparticle pole dominates 
the dynamics and the contribution of the continuum becomes subleading,
and the relaxation time scale which is determined 
by the imaginary part of the self energy at the quasiparticle pole,
i.e. the width of the resonance. In this case, the two different
time scales can only be resolved if they are widely separated which
requires that the resonance be very 
narrow and the exponential relaxation associated with the decay of the
quasiparticle allows many oscillations to occur. 
This condition can be quantified as $\Gamma / \omega_p<<1$ which
requires a weak coupling to the bath. We will explore 
these situations numerically in a later section where a particular
density of states of the bath will be proposed.

\subsection{Asymptotic behavior of $n(t)$}

The asymptotic  behavior of $\left<n(t)\right>$ is completely
determined by the long time dynamics   
of $g(t)$. 
We have shown that $g(t)$ vanishes asymptotically for poles in the
continuum while the contribution from the isolated pole  dominates for
the case in which the pole is below  threshold. We will consider each
individual case in detail.\newline  

 \textbf{I)} $\omega_p > \omega_{th} $ : in this case the function
$g(t)$ vanishes exponentially at asymptotically long times
(\ref{abovecut}) and the asymptotic behavior of the particle  
occupation number is given by
\begin{equation}
\left<n(\infty)\right> = -\frac{1}{2} + {\cal R}(\infty)
\label{nasin}
\end{equation}
with
\begin{equation}
{\cal R}(\infty) = \frac{1}{2\pi \Omega} \int^{\omega_c}_{\omega_{th}}
 d\omega \; \left[1+2N(\omega)\right]\; 
 S(\omega) \;\left(\Omega^2 + \omega^2 \right)\;,\label{Rinfty}
\end{equation}
where we used eqs.(\ref{imsig}) and (\ref{specdef}), 
recognized the Laplace transform of $g(t)$ in the long time
limit for eq.(\ref{siguiente})
(using the vanishing of $g(t)$ at long times)
\begin{eqnarray}
\left|h(\omega,\infty)\right|^2 & = & \left| \tilde{g}(s=i\omega +
\epsilon)\right|^2 
\nonumber\\
\left|k(\omega,\infty)\right|^2 & = & \omega^2
\left|h(\omega,\infty)\right|^2\; \; .\nonumber
\end{eqnarray} 
It is clear that the asymptotic value of $\left<n(\infty)\right>$ is
different from the  
equilibrium occupation number of the bath $N(\omega_p)$.


Suppose that the spectral density $S(\omega)$ can be approximated by a narrow
Breit-Wigner resonance with
\begin{equation}
S(\omega)\,=\,\frac{Z}{2\omega_p }~ \frac{\Gamma}{(\omega - \omega_p)^2 + \Gamma^2}
\stackrel{\Gamma \rightarrow 0}{\rightarrow} \quad \frac{\pi Z}{2\omega_p } \delta(\omega-\omega_p),
\end{equation}
where
\begin{equation}
\Gamma = \frac{Z \Sigma_I(\omega_p)}{2 \omega_p}
\label{gamma}
\end{equation}
as would be the case for weak coupling.
Then the asymptotic occupation number becomes

\begin{equation}
\left<n(\infty)\right> = {Z}\left( 
\frac{\Omega^2+\omega^2_p}{2\Omega \omega_p} \right) \left[N(\omega_p)
+ \frac{1}{2}\right] - \frac{1}{2}\;, 
\label{nasab}
\end{equation}
which is different from the equilibrium value of the bath.
We now see that choosing the reference frequency $\Omega=\omega_p$, 
the particle thermalizes
completely with the bath, i.e.  asymptotically the occupation number
becomes the one predicted by the canonical ensemble as we will see
below in eq.(\ref{thermo}). 
This expression, thus reveals the importance of counting the
quasiparticles instead of the bare particles.  Even in the
weak coupling limit the distribution of  
bare particles is not  thermal whereas the true
quasiparticles are described by a  thermal distribution (that differs from
that of Bose-Einstein by perturbatively small terms).  

The asymptotic value of the distribution is approached
exponentially. The thermalization time scale is given by $\tau_{th} =
1/2\Gamma$ since 
it is determined by $g^2(t)$ which is the dependence of the occupation
number on the function that determines the real time evolution either 
of the density matrix or of the Heisenberg operators.

Even when the occupation number is defined in terms of the
true `in medium' pole, there will be departures from the Bose-Einstein
distribution for non-negligible width $\Gamma$ and when the strength
of the pole $Z$ is substantially smaller than one. These corrections
will arise in the case of broad resonances and may lead to large
departures from the Bose-Einstein distribution. This situation will 
be explored numerically later.  

In the case of a wide resonance, the product $N(\omega)\, S(\omega)$ is
sensitive to the width of the resonance.  
For bath temperature $ T << \omega_{th} $ the Bose-Einstein
distribution will only probe the tail of the broad spectral 
density closer to threshold and the product is only sensitive to the
threshold behavior of $S(\omega)$\cite{yoshimura}. 

In particular if near threshold  $S(\omega) \approx
(\omega-\omega_{th})^{\alpha}$ then for temperatures $T <<
\omega_{th}$ 
the temperature dependence of the equilibrium abundance of unstable
particles in the bath is approximately given by 
$$
n(T;t= \infty)-n(0;t= \infty) \approx e^{-\frac{\omega_{th}}{T}}~
T^{\alpha+1} 
$$

\noindent which reveals threshold corrections to the Boltzmann exponential 
suppression. This  result has been  anticipated in\cite{yoshimura}
within a different context.  

In the opposite limit when $T >> \omega_p$  the product is sensitive
to the width of the resonance and the details 
of the spectral density. Thus in the case of a broad resonance the
departures from a Bose-Einstein distribution function for the
quasiparticles will be  
important. Clearly this is the regime in which a Boltzmann
approximation could be unreliable.  

These corrections originate in  off-shell
effects that will depend on the particular spectral density of the bath
and the coupling between the particle of the bath. We will quantify the
corrections for a particular choice of $J(\omega)$ in a following section.

 Moreover, the asymptotic
value of the particle number does not depend on the initial condition of the particle; e.g.
initial expectation values of position and momentum, temperature or occupation number.

\noindent  \textbf{II)} $\omega_p < \omega_{th} $ :
In this case the asymptotic time dependence of the function $g(t)$ is
completely determined by the isolated pole below the continuum  and
the function rings with this frequency with asymptotic amplitude
determined by the wave function renormalization 
$Z$ given by (\ref{wavefunc}). The asymptotic behavior of the 
 particle occupation number defined at a reference frequency $\Omega$
is now given by 
\begin{eqnarray}
\left<n(\infty)\right> & = & -\frac{1}{2} \,+\, {\cal R}(\infty)
\,+\,\frac{Z^2 \sin^2(\omega_p t)}{2\,\Omega \omega_p^2 }
\left[p_i^2 \Omega^2+ ( \Omega^4 + \omega_p^4) \sigma +  \omega_p^4
q_i^2 \right]
\nonumber \\
& &+ \; \frac{p_i q_i Z^2}{2 } \left(\frac{\Omega}{\omega_p} -
\frac{\omega_p}{\Omega}\right) 
\,\sin(2 \omega_p t)\,+\, 
\frac{Z^2\cos^2(\omega_p t)}{2}
\left(\frac{p_i^2}{ \Omega} + 2  \Omega \sigma +  \Omega q_i^2
\right), \label{nasy1} 
\end{eqnarray}
where ${\cal R}(\infty)$ is the limit value of ${\cal R}(t)$. 
For $\Omega = \omega_p$, i.e. the position of the dressed particle
pole, the asymptotic value of the occupation number obtains  the
simple form 
\begin{equation}
\left<n(\infty)\right>  =  -\frac{1}{2} \,+\, {\cal R}(\infty) + {Z^2} \left[
n(0) + \frac{1}{2} \right] + \frac{Z^2}{2 \Omega}\left[ 
{p_i^2}+  \Omega^2 q_i^2\right].
\label{nasb1}
\end{equation}

The last term can be identified as the contribution from the expectation values of $p(0)\; ; \;q(0)$ in the initial density matrix.  

Unlike the case in which the pole is in the continuum, the asymptotic
value of the particle  
occupation does depend on how the particle was prepared initially since 
expression (\ref{nasb1}) depends on $p_i$, $q_i$ and $n(0)$.

In this case, ${\cal R}(\infty)$ 
has contributions from both the continuum  cut and the isolated pole
below the continuum.

In order to compare the results to those obtained from an {\em
approximate} quantum kinetic equation obtained via a perturbative  
expansion in the next section,  it is 
useful to obtain an expression for ${\cal R}(\infty)$ up to first
order in $J(\omega)$. The expression for   
${\cal R}(\infty)$ (\ref{Rinfty}) is proportional to the spectral density
$S(\omega)$ given by (\ref{specdef}). When the pole is below the
continuum, the contribution from the cut is proportional to
$J(\omega)$ and perturbatively small when $J(\omega)$ is small.  
Furthermore the continuum contribution dephases rapidly at long times,
 and asymptotically the relevant contribution to $g(t)$ arises from
the isolated pole. After some straightforward algebra we find for
$\Omega=\omega_p$ that at long times  
$$
|k(\omega,t)|^{2}+ \Omega|h(\omega,t)|^{2}\;=\;Z^2 \big \{
\frac{1-\cos(\omega_{+}t)}{\omega_{+}^{2}}+
\frac{1-\cos(\omega_{-}t)}{\omega_{-}^{2}}\big \}  
$$
with 
\begin{equation}
\omega_{\pm} \equiv \Omega \pm \omega 
\label{wpm}
\end{equation}
and to lowest order in $J(\omega)$, the asymptotic contribution ${\cal
R}(\infty)$ is given by   
$$
{\cal R}(\infty) \,=\, \frac{Z^2}{2 \pi  \Omega}\int d\omega \; J(\omega)\;
\left[1+2N(\omega)\right]\left(\frac{1}{\omega_{+}^{2}}+
\frac{1}{\omega_{-}^{2}}\right)+ {\cal O}\left(J^2\right)\; ,
$$
 which for easier comparison with the results from kinetics, can be
 written in the following form  
$$
{\cal R}(\infty) \,\approx \, \frac{1}{2}(1-Z^2) + \frac{Z^2}{\pi  \Omega}
\int d\omega\;  J(\omega) \left\{ \frac{1+N(\omega)}{\omega^2_+} + 
\frac{N(\omega)}{\omega^2_-} \right\} + {\cal O}\left(J^2\right)\; ,
$$
where the term $(1-Z^2)\approx 2(1-Z)$ and we have used the sum
rule (\ref{sumrulebelowcut}) to lowest order.

Setting $p_i=q_i=0$ in \mbox{eq.(\ref{nasb1})}, the asymptotic
occupation number becomes 
\begin{eqnarray}
\left<n(\infty)\right> &=& Z^2 \left[ n(0)+ 
\frac{1}{\pi  \Omega} \int d\omega J(\omega) \left\{
\dfrac{1+N(\omega)}{\omega^2_+} + 
\dfrac{N(\omega)}{\omega^2_-} \right\}
\right] + {\cal O}  \left( J^2 \right) \; .  
\label{exres}
\end{eqnarray}
\noindent We have purposedly kept $Z$ in the above expression to compare
it to the results from the quantum kinetics approximation to be obtained
later. 

Clearly this result depends on the initial distribution of the particle 
and the details of the spectral density of the bath, leading to the
conclusion that in the case in which the particle pole is real 
(below threshold), the particle {\bf does not thermalize} with the bath. 
%
%
%

\section{Collective Normal Modes and Quasiparticles \label{secnorm}}

In a many body problem, the poles of the exact two particle Green's
functions are identified with the collective modes.
 In general the poles are complex resulting in the damping of the collective 
excitations. We can make contact with this many body concept by studying
the {\em normal modes} of the total Hamiltonian for the particle-bath
system under consideration. 

Since the Hamiltonian is quadratic, it can be diagonalized by a
canonical transformation in terms of the normal 
modes. In order to establish a correspondence with the continuum
distribution of bath oscillators it is convenient to 
 write the Hamiltonian in the continuum form

\begin{eqnarray}
H & = &  \frac{1}{2}(p^2 + \omega^2_0 ~ q^2) +
\frac{1}{2}\int_{\omega_{th}}^{\omega_c} d\omega 
\left[ P^2(\omega) + \omega^2 \; Q^2(\omega)\right] + q
\int_{\omega_{th}}^{\omega_c} d\omega \; 
C(\omega)\;  Q(\omega) \nonumber  \\
J(\omega) & = & \pi\;  \frac{C^2(\omega)}{2\omega}\; . \nonumber 
\end{eqnarray}

The Hamiltonian of this rather simple model can be diagonalized by
finding the normal modes. Let us write the linear change coordinates
and momenta (canonical transformation) to the normal modes
as\cite{ullersma,yoshimura} 
\begin{eqnarray}
q & = &  {\bf{\cal S}}_{\lambda}~ \alpha(\lambda) {\cal Q}(\lambda)
\quad ; \quad  
p = {\bf{\cal S}}_{\lambda}~ \alpha(\lambda) {\cal P}(\lambda)
\label{littleqp} \\  
Q(\omega) & = & {\bf{\cal S}}_{\lambda}~\beta(\omega,\lambda) {\cal
Q}(\lambda) \quad ; \quad 
P(\omega) = {\bf{\cal S}}_{\lambda}~ \beta(\omega,\lambda) {\cal
P}(\lambda), \label{bigqp} 
\end{eqnarray}
where the symbol ${\bf{\cal S}}_{\lambda}$ stands for the sum over the
 discrete and integral over the continuum normal mode 
 eigenvalues $\lambda$  that render the Hamiltonian in diagonal form 
$$
H= \frac12 {\bf{\cal S}}_{\lambda}~\left[ {\cal P}^2(\lambda)+\lambda^2 {\cal
 Q}^2(\lambda)\right]\;.  
$$
The vectors $V(\lambda) =
\left(\alpha(\lambda),\beta(\omega,\lambda)\right)$ obey the normal
mode eigenvalue equation which in components reads 
\begin{eqnarray}
\omega^2_0 \; \alpha(\lambda) + \int_{\omega_{th}}^{\omega_c} d \omega\; 
C(\omega)\;  \beta(\omega,\lambda) & = & \lambda^2 \; \alpha(\lambda)
\label{normal0} \\ 
C(\omega) \; \alpha(\lambda)+ \omega^2 \beta(\omega,\lambda) & = &
\lambda^2 \; \beta(\omega,\lambda) 
\label{normalk} 
\end{eqnarray}
and the $\lambda's$ are the {\em exact} eigenenergies of the Hamiltonian.

 Solving for $\beta(\omega,\lambda)$ in terms of $\alpha(\lambda)$ in
eq.(\ref{normalk}) and inserting the solution back into (\ref{normal0}) we
find the solution for the coefficients and the secular equation for
the eigenvalues to be given by 
\begin{eqnarray}
& & \beta(\omega,\lambda)  =
\frac{C(\omega)\; \alpha(\lambda)}{(\lambda-i\epsilon)^2-\omega^2}+ B\;
\delta(\lambda-\omega) 
\nonumber \\
&  & \left[\lambda^2-\omega^2_0 -\frac{2}{\pi}
\int_{\omega_{th}}^{\omega_c} d\omega  
\frac{\omega \; J(\omega)}{(\lambda-i\epsilon)^2-\omega^2}
\right]\alpha(\lambda) = B \; C(\lambda)\;,  \nonumber
\end{eqnarray}
where we used `retarded' boundary conditions (with the $i\epsilon$
prescription) to establish contact with the previous results,  and $B$
is  determined by normalizing the eigenstates.

There are two distinct possibilities: I) an isolated pole below the
continuum threshold of the bath corresponding to 
a dressed stable particle, II) a continuum
of states and a quasiparticle pole in the unphysical Riemann sheet
(resonance).  

{\bf I) Isolated poles:} The condition for isolated poles below the
bath continuum requires setting $B=0$ since the 
spectrum of the bath has no support below threshold. The position of
the pole is found from the secular equation 
$$
\omega_p^2-\omega^2_0 -\frac{2}{\pi} \int_{\omega_{th}}^{\omega_c} d\omega 
\frac{\omega \, J(\omega)}{\omega_p^2-\omega^2}= 0\; .
$$
This expression is identified as the condition for isolated poles in
the Laplace transform $\tilde{g}(s)$ (see eq.(\ref{glap})) for $s=i\omega_p$.
The value of $\alpha(\omega_p)$ is determined from normalization and we find
$$
\alpha(\omega_p) = \sqrt{Z} 
$$

\noindent with $Z$ the wave function renormalization given by
eqs.(\ref{realsig}) and (\ref{wavefunc}). Normalization of the vectors is
equivalent to the sum rule (\ref{sumrulebelowcut}).

{\bf II) Continuum states:} 

For the continuum states we take $B=1$ so that ${\cal Q}(\lambda) \rightarrow Q(\omega)$ when $C(\omega) \rightarrow 0$
and we find the coefficients
\begin{eqnarray}
&& \displaystyle{\alpha(\lambda) =
\frac{C(\lambda)}{(\lambda-i\epsilon)^2-\omega^2_0 -\frac{2}{\pi}
\int_{\omega_{th}}^{\omega_c} d\omega  \;
\frac{\omega\; J(\omega)}{(\lambda-i\epsilon)^2-\omega^2}}}
\nonumber \\ 
&& \beta(\omega,\lambda) = \delta(\lambda-\omega) +
 \frac{C(\omega)\;
 \alpha(\lambda)}{(\lambda-i\epsilon)^2-\omega^2}\;.\nonumber
\end{eqnarray}
In this case the normalization results in the sum rule given by
eq.(\ref{sumruleabovecut}). 

Because of our choice of boundary conditions, the coefficients are
complex and the resulting new coordinates are not 
Hermitian, this can be remedied by absorbing the phases by a trivial
canonical transformation and defining the coefficients 
in terms of their absolute values. This phase carries the information of the boundary conditions
(the $i\epsilon$ prescription) and since it is removed by a canonical transformation the results are independent of
these. 

Let us consider the case of an isolated pole below the threshold of the
bath continuum at $\lambda = \omega_p$. This state is the one that
evolves from the bare particle degree of freedom upon adiabatically
switching-on the system-bath couplings $C_k$ and is identified
with the position of the isolated pole in the Laplace transform of the
function $g(t)$ given by (\ref{glap}).

Separating the contribution from the isolated pole  we  write 
\begin{eqnarray}
q(t) & = &  \sqrt{Z} {\cal Q}_0(t) + {\cal Q}_{\mbox{cont}}(t)
 \nonumber  \\ 
 p(t) & = &  \sqrt{Z} {\cal P}_0(t) + {\cal P}_{\mbox{cont}}(t)\; ,
 \nonumber 
\end{eqnarray}
where the operators ${\cal Q}_{\mbox{cont}}\; ; \; {\cal
P}_{\mbox{cont}}$ create excitations in the continuum of 
the bath out of the {\em exact} ground state. Writing ${\cal Q}_0 \; ;
\; {\cal P}_0$ in terms of creation and annihilation operators of 
the {\em exact} eigenstates, we see that asymptotically long times the
operators $q(t)\; ; \;  
p(t)$ create an {\em exact} one    {\em dressed particle} state out of the {\em exact} vacuum.
In the limit of asymptotically long times and invoking the
Riemann-Lebesgue lemma 
$$
q(t) {|\bf{0}>} \; \; {\rightarrow}\; \; 
\frac{\sqrt{Z}}{\sqrt{2\omega_p}}e^{i\omega_p t}\;  {|\bf{1}_p>}, 
$$
where ${|\bf{0}>}$ is the {\em exact} ground state and  the contribution from the continuum states averages to zero at long
times by the dephasing between modes.

The operator
\begin{equation}
A^{\dagger}_{q}(t) = \frac{1}{\sqrt{2\omega_p Z}}
\big[\omega_p\; q(t) + i p(t)\big] 
\label{quasicreation}
\end{equation}

\noindent asymptotically at long times creates a dressed particle
state with unit residue out of the exact vacuum. At any finite time 
the state created by this operator is {\em not} an eigenstate of the full
Hamiltonian but has overlap with states in the continuum. We associate
the operator (\ref{quasicreation}) with dressed particles in the case of isolated poles or
{\em quasiparticles} for resonances, in contrast
to the normal (collective) modes of the system that are exact eigenstates.

Although a priori one would be tempted to define the dressed particle as
the normal mode of frequency $\omega_p$ associated 
with the creation and annihilation operators obtained from the normal
mode described by ${\cal Q}_0 \; ; \; {\cal P}_0$, these are of little
use: these operators represent linear combinations of the particle and
the degrees of freedom of the bath. Obviously the number  operator
associated with this normal mode is constant in time. Furthermore, in the case of a resonance, there is no
operator that creates or destroys a quasiparticle and  that can be understood as the limit of a bare particle operator upon adiabatic switching on of the interaction. Thus the interpolating operator (\ref{quasicreation}) is the natural
candidate for counting. 

In an experimental situation such as for example a lepton in a plasma,
or an electron in a metal, one would like to 
write down an evolution equation for the distribution  function that
describes the particle dressed by the medium 
effects. The interpretation of the quasiparticle creation operator
is consistent with this physical situation since 
the added particle will move in
the bath being dressed by the interaction with the medium, the resulting
quasiparticle will have a new dispersion relation (given here by $\omega_p$) and in general a width,  and the probability associated with this quasiparticle pole will be reduced
by the overlap with the states of the bath.  Whereas the collective modes in the bath are stationary states, this quasiparticle is not because it
overlaps with the collective modes and its time evolution involves dephasing.  

In the case in which the pole at $\omega_p$ has a value larger than
the threshold for the bath oscillators, it has moved into the second (unphysical) Riemann sheet upon adiabatically switching-on the interaction and is no longer part of the eigenspectrum of the Hamiltonian. In this case it has become an unstable state and $\omega_p$ will have a small negative imaginary part given by $\Gamma$ [see eq.(\ref{abovecut})].  In this case the
overlap with the continuum results in an almost exponential decay of
the quasiparticle distribution after the short formation time of the
quasiparticle.

We then see that the {\em interpolating} number operator 
\begin{equation}
\hat{n}_{quasi}(t) = \frac{1}{2\omega_p Z} \left[p^2(t)+\omega^2_p
q^2(t)\right] -\frac{1}{2Z}\label{quasi} 
\end{equation} 
\noindent can be interpreted as either the dressed particle
distribution function in the case of an isolated pole below the
continuum of the bath or the quasiparticle distribution function in
the case of a resonance. Besides setting the reference frequency
$\lambda \equiv \omega_p$ in eq.(\ref{number}) 
the wavefunction renormalization factor $Z$ accounts for the strength
of the particle or quasiparticle pole.

{\bf Interpretation of results:}

This analysis in terms of normal modes reveals several features of the
exact solutions obtained in the previous sections.  

\begin{itemize}

\item {{\bf Thermalization of resonances:} in the case in which the
quasiparticle pole is above threshold, the asymptotic value of the 
quasiparticle distribution given by eqs.(\ref{nasin})-(\ref{Rinfty}) is
a consequence of {\em thermalization}. Indeed by using the expansion of
$q \; ; \; p$ in terms of the normal mode coordinates and momenta
given by eqs.(\ref{littleqp})-(\ref{bigqp}) it is 
straightforward to prove that
\begin{equation}
<\hat{n}_{quasi}(\infty)> = Tr\left[\hat{n}_{quasi}(0)~ e^{-\beta
H}\right]  \label{thermo} 
\end{equation}

with $\hat{n}_{quasi}(0)$ the quasiparticle number operator
(\ref{quasi}) at the 
initial time $t=0$. This is
a remarkable result: the density matrix, which initially was of a
factorized form for particle and bath at different temperatures has 
evolved in time to the {\em equilibrium} density matrix for the total
system at the temperature of the bath. However the distribution of 
quasiparticles is {\bf not} given by the Bose-Einstein
form. Furthermore, the contribution to $\hat{n}_{quasi}(\infty)$ that
does not vanish 
as $T \rightarrow 0$ can be interpreted as a zero point contribution
from the resonance. In the case in which the quasiparticle becomes a
narrow resonance  we see 
from eqs.(\ref{nasab}) and (\ref{quasi}) that 
\begin{equation}
<n_{quasi}(\infty) > = N(\omega_p) +\frac{1}{2}(1-\frac{1}{Z})\label{asynumi}
\end{equation} 
and the number of quasiparticles departs  from a Bose-Einstein
distribution at the temperature of the bath with the departure
determined by the  off-shell effects that result in $Z \neq 1$ through
the sum rules. The last term, identified above with the zero point
contribution 
is interpreted as the normalization borrowed from the continuum by
the quasiparticle. Although in this simple case $Z$ does not depend
on temperature and the last term in (\ref{asynumi}) can be subtracted out
as a redefinition of the quasiparticle vacuum, in a general field
theory, the wave function renormalization will be medium dependent and
such subtraction would be unjustified.} 

\item {{\bf Non-Thermalization of stable particles:}
In the case in which the particle pole is below threshold the
asymptotic oscillations in the expression (\ref{nasy1}) for $\Omega
\neq \omega_p$ are a  
consequence of the interference between the state of arbitrary
frequency $\Omega$ and the normal mode with frequency
$\omega_p$. These oscillations disappear when the reference frequency
($\Omega$) is chosen to be the normal mode pole frequency ($\omega_p$)
which is also the particle frequency, this fact has already been
noticed within a 
different context\cite{boyprem}. The factor $Z^2$ in eq.(\ref{nasb1}) has
the following origin: asymptotically at long times  
$q(t) \rightarrow \sqrt{Z} {\cal Q}_0 \; ; \;  p(t) \rightarrow
\sqrt{Z} {\cal P}_0$ in the sense of matrix elements. 
But the ${\cal Q}_0 \; ; \; {\cal P}_0$ create particle states
out of bare states with amplitude $\sqrt{Z}$, therefore 
one of the factors $Z$ in eq.(\ref{nasb1}) arises from the asymptotic
(weak) limit on the operators, and another factor 
${Z}$ arises because the  calculation of eq.(\ref{nasb1}) was performed
in terms of the bare states overlap with the 
 particle states given by the wave function renormalization.
Using the expansion in terms of normal modes we find that 
  $<\hat{n}_{quasi}(\infty)>$ given by eq.(\ref{quasi}) does {\bf not}
coincide with 
$$
Tr\left[\hat{n}_{quasi}(0)~e^{-\beta H}\right]
$$ 
unlike the previous case of a resonance.}

\end{itemize}

%
%
%
%
%
%
%
%

\section{Kinetics \label{seckin}}
Having provided an analysis of the exact evolution of the distribution
function and distinguished between that of particles and quasiparticles,
we now proceed to obtain kinetic equations in several stages of
approximation to compare with the exact results. Kinetic equations are
obtained by truncating the hierarchy of Schwinger-Dyson equations under
certain assumptions. The typical assumptions are those of slow
relaxation as compared to the microscopic time and length scales and
rely on a 
separation of scales, or a multi-time scale problem. To warrant this
separation between scales clearly a perturbative parameter must be
invoked and the resulting kinetic equations provide a resummation of 
the perturbative expansion. Different type of approximations result in
different resummation schemes.

\subsection{Boltzmann Equation}
The simplest kinetic equation to describe the approach to equilibrium
is the Boltzmann equation, which is obtained by writing a gain minus loss
balance equation in which energy is conserved and using Fermi's Golden Rule.
Writing $q(t)$ and $Q_k(t)$ in terms of creation and annihilation
operators, we identify the only terms that can contribute 
by energy conservation: the `gain' term in which one quanta of the
system's oscillator is created and one quanta of the oscillator with
label $k$ is annihilated, minus the `loss' term  in which a quanta
of the system's oscillator is annihilated and a quanta of the
oscillator of label $k$ in the bath is created. The first term has
probability given by 
$$
\mbox{gain} = \frac{C^2_k}{4\omega_0~ \omega_k} (1+n)N_k.
$$
The second term has a probability
$$
\mbox{loss} = \frac{C^2_k}{4\omega_0~ \omega_k} (1+N_k)~n.
$$
Thus using Fermi's Golden Rule we find the usual Boltzmann equation
\begin{eqnarray}
\left<\dot{n}_{B}(t)\right> & = &  (2 \pi) \sum_k \frac{C^2_k}{4 \omega_0 \omega_k}\left[
(1+n(t))N_k - (1+N_k)n(t) \right] \delta(\omega_k-\omega_0) \rightarrow \nonumber \\
&  & \int d \omega ~ \frac{J(\omega)}{\omega_0} \left[(1+n(t))N(\omega)- (1+N(\omega))n(t) \right] \delta(\omega-\omega_0), \label{Boltzmann}
\end{eqnarray}
where we have assumed that the bath remains in thermal equilibrium with
constant distribution functions. The  solution is clearly
\begin{equation}
\left<n_B(t)\right> = N(\omega_0)+\left[n(0)-N(\omega_0)\right] e^{-2\Gamma_B t} \; ; \; 
\Gamma_B= \frac{J(\omega_0)}{2\omega_0} = \frac{\Sigma_I(\omega_0)}{2\omega_0}, 
\label{eqboltz}
\end{equation}

\noindent where we recognize the lowest order (Born approximation) to
the decay rate 
which is given by eqs.(\ref{imsig}) and (\ref{abovecut}). Obviously the
Boltzmann equation predicts no relaxation in the case in which the  pole
is below the continuum threshold since in this case $J(\omega_0)=0$.

Even when the bare frequency is in the continuum of the bath, the
Boltzmann approximation predicts no relaxation if $n(0)= N(\omega_0)$
as the gain and loss processes exactly balance.  
 
As we will see explicitly numerically below the exact solution shows
a non-trivial time dependence in this case because the bare particle
is dressed by the medium and the asymptotic equilibrium distribution
function reveals off-shell effects as discussed in the previous
section.  

\subsection{Quantum Kinetic Equation:}
The quantum kinetic equation is obtained by taking the expectation
value of the number operator using the Heisenberg equations of motion
and truncating the exact equations of motion within a particular
approximation.  

Since we want to obtain the kinetic equation for the relaxation of the
distribution function of particles with frequency $\Omega$ (for
quasiparticles this is the pole frequency of the propagator, for
bare particles it is simply $\omega_0$) it is convenient to write the
total 
Hamiltonian in terms of this frequency adding a counterterm of the
form
$$
H_{ct} = \frac{\delta\omega^2}{2} \; q^2(t) \quad ; \quad \delta\omega^2 =
\omega^2_0 - \Omega^2\; . 
$$
As usual the counterterm is chosen appropriately in perturbation theory
to cancel the contributions recognized as those arising from a shift
in the frequency.

Taking the derivative of eq.(\ref{number}) and using the equations of 
motion we obtain
\begin{equation}
\dot{n}(t) = -\frac{1}{\Omega} \left\{ \sum_k {C_k}\; Q_k(t) \dot{q}(t)+
\frac{\delta \omega^2}{2}
\left[q(t)\;\dot{q}(t)+\dot{q}(t)\;q(t)\right] \right\}\;. 
\label{ndot}
\end{equation}
The expectation value of the time derivative of the occupation number is calculated
by multiplying eq.(\ref{ndot}) by $\rho(0)$ and taking the trace 
\begin{equation}
\left<\dot{n}(t)\right> = -\frac{1}{\Omega}\frac{d}{dt'}\left. \left\{
 \sum_k C_k 
 \left<q^+(t')\,Q^-_k(t)\right>
+ \frac{\delta \omega^2}{2} \left<q(t)q(t')+q(t')q(t)\right> \right\}
 \right|_{t'=t},
\label{ndot1}
\end{equation}
where
$$
\left<q^+(t')\,Q^-_k(t)\right> = \mbox{Tr}\;[ {q}(t') \rho(0) Q_k(t)
]\; . 
$$
We need to evaluate the non-equilibrium matrix element
$\left<q^+(t')\,Q^-_k(t)\right>$. 
This can be achieved by treating the interaction term in perturbation theory. The zeroth order
term in the perturbative series does not contribute because the initial density matrix commutes with the number operator
at the initial time.

A simple diagrammatic analysis of the perturbative series reveals that the
kinetic equation can be written {\em exactly} as
\begin{eqnarray}
\left<\dot{n}(t)\right> & = & -\frac{1}{\Omega} \sum_k {C_k}
\frac{d}{dt'} \left\{  
\left. \left[\int_0^t dt''\left( \Sigma^<_k(t,t'') {\cal
G}^>(t'',t')-\Sigma^>_k(t,t'') {\cal G}^<(t'',t')\right) \right]
+\right. \right. \nonumber \\ 
&  & \left. \left. \frac{\delta \omega^2}{2} \left({\cal G}^>(t,t')+
{\cal G}^<(t,t')\right) \right\}\right|_{t'=t}\; , \nonumber
\end{eqnarray}
where ${\cal G}^{<,>}$ are the {\em exact} Green's functions for the
system, defined 
similarly to those of the bath eq.(\ref{greens}) and $\Sigma_k^{<,>}$ are
the irreducible self-energy components, again defined similarly to
eq.(\ref{greens}).

 To first order in
the interaction we use the free-field propagators and the lowest order
contribution to the self-energy. It is 
straightforward to show that the counterterm contribution vanishes to
this order and  \mbox{eq.(\ref{ndot1})} becomes 
\begin{eqnarray}
\left<\dot{n}(t)\right> &=& \frac{i}{\Omega} \sum_k {C^2_k}
\frac{d}{dt'} \left[ \int_0^\infty dt''\left(
\left<q^+(t')\,q^+(t'')\right>\left<Q^-_k(t)\,Q^+_k(t'')\right> 
\right. \right.\nonumber \\
& & \hspace{1.5in} \,-\left.\left.
\,\left<q^+(t')\,q^-(t'')\right>\left<Q^-_k(t)\,Q^-_k(t'')\right>
\right)\right]_{t'=t}\; . 
\label{ndot2}
\end{eqnarray}
Substituting the non-equilibrium Green's functions from
eq.(\ref{greens}) in the right 
hand side of eq.(\ref{ndot2}), taking the derivative with respect to
$t'$ and arranging terms, we obtain
\begin{eqnarray}
\left<\dot{n}(t)\right> &=& \frac{1}{\pi  \Omega} \int_0^t dt'
\int d\omega J(\omega)\left\{
\left[1+n(0)+N(\omega)\right]\cos\left[(\Omega+\omega)(t-t')\right]
\right.\nonumber \\ & & \hspace{1.5in} \,+\left.
\left[N(\omega)-n(0)\right]\cos\left[(\Omega-\omega)(t-t')\right] 
\right\},
\label{bkin}
\end{eqnarray}
where $n(0)$ is the distribution of quanta for the particle {\em at
the initial time} and $N(\omega)$ are the Bose-Einstein distributions
of the bath which will be taken to be constant and given by
eq.(\ref{Nk}).  

We now propose a scheme that provides a resummation of the
perturbative series by replacing the initial distribution $n(0)$ by
self-consistently 
updating the distribution  inside the integral in  
eq.(\ref{bkin}) by replacing $n(0) \rightarrow n(t')$. It will be shown
explicitly below that this prescription leads to a  Dyson summation of
particular Feynman diagrams and the case where $n$ is constant is
understood as the lowest order term in this expansion. Within 
non-relativistic many-body quantum kinetics, this approximation is
known as the generalized Kadanoff-Baym ansatz\cite{meden,greiner}. The
validity of this approximation in the weak coupling limit is confirmed
by comparing the 
resulting evolution of the distribution function to the exact result
obtained in the previous sections.

This resummation scheme has been recently invoked to study relaxation in the case of gauge theories in the hard-thermal loop limit\cite{boyprem} and to provide a resummation scheme that incorporates non-perturbatively the effects of
instabilities   in the photon production mechanism during the chiral
phase transition \cite{boyphoton}.  
 The quantum kinetic equation is then given by
\begin{eqnarray}
\left<\dot{n}_{qk}(t)\right> &=& \frac{1}{\pi  \Omega} \int_0^t dt'
\int d\omega \; J(\omega)\left\{
\left[1+n(t')+N(\omega)\right]\cos\left[(\Omega+\omega)(t-t')\right]
\right.\nonumber \\ & & \hspace{1.5in}
\,+\left. \left[N(\omega)-n(t')\right]\cos\left[(\Omega-\omega)(t-t')\right] 
\right\}. \label{kin}
\end{eqnarray}

The resulting linear kinetic equation, eq.(\ref{kin}), can now be
solved via  Laplace transforms. The Laplace transform of $<n_{qk}(t)>$
is given by  
\begin{eqnarray}
\tilde{n}_{qk}(s) &=& \frac{
n(0)+ \frac{1}{\pi  \Omega} \int^{\omega_c}_{\omega_{th}} d\omega J(\omega) \left\{
\frac{\left(1+N(\omega)\right)}{s} \frac{s}{s^2+\omega^2_+} + 
\frac{N(\omega)}{s} \frac{s}{s^2+\omega^2_-} \right\}}{
s - \frac{1}{\pi  \Omega}\int^{\omega_c}_{\omega_{th}} d\omega J(\omega) \left\{\frac{s}{
s^2+\omega^2_+} - \frac{s}{s^2+\omega^2_-} \right\}},
\label{nkinlap}
\end{eqnarray}
where $n(0)$ is the initial occupation number of the particle and
$\omega_\pm$ are given 
by eq.(\ref{wpm}). The dynamics of the occupation number of the
particle is obtained by 
taking the inverse Laplace transform along the Bromwich contour. The
analytic structure 
of $\tilde{n}_{qk}(s)$ consists of cuts along the imaginary axis in
the $s$-plane and  
poles. For the pole contributions, we distinguish two cases :\newline
{\textbf Case I} : Poles in the continuum. In this case there are two
poles: 1) a pole 
where the denominator of eq.(\ref{nkinlap}) vanishes, i.e.
$$
s_p - \dfrac{1}{\pi  \Omega}\int d\omega J(\omega) \left\{\frac{s_p}{
s_p^2+\omega^2_+} - \frac{s_p}{s_p^2+\omega^2_-} \right\} = 0.
$$
For weak coupling, one can solve for the pole, $s_p$, in perturbation 
theory and one can show that the pole is given by (up to first order
in $J(\omega)$) 
$$
s_p = -\frac{J(\Omega)}{ \Omega} = -2 \Gamma,
$$
where we used the fact
\begin{equation}
\lim_{s\rightarrow 0}\frac{s}{s^2+\omega^2_\pm} = \pi \delta(\omega_\pm),
\label{del1}
\end{equation}
and $\Gamma$ is given by eq.(\ref{gamma}). The contribution from this
pole vanishes  
exponentially for long times. 
2) There is a second pole at $s=0$ and the residue of this pole, using eq. 
(\ref{del1}), is $N(\Omega)$. The average occupation number is then
given by the contribution of the two poles and the cut 
$$
\left<n(t)\right> = N(\Omega) + \mbox{(residue at $s_p$)}\; e^{-2\Gamma\,
t} + \mbox{ contribution from the cut}\; . 
$$
Asymptotically, the contribution from the last two terms vanish and
the particle occupation 
number goes to the equilibrium occupation of the bath with frequency
$\Omega$. Comparing the 
above result for $\Omega = \omega_p$ with the one obtained exactly in
the small coupling regime, \mbox{eq.  
(\ref{nasab})}, we see that they differ by a factor of order
$J(\omega)$ which can 
be compensated for by considering higher orders in deriving the
kinetic equation. Thus we see 
that for weak coupling ($\omega_p \approx \omega_0$), the solution of
the quantum kinetic  
equation approaches that of the Boltzmann 
approximation given by eq.(\ref{eqboltz}).

\noindent {\textbf Case II} : Poles below the continuum. Since $J(\omega_p)$ 
vanishes for poles
below the continuum, there is only one pole at $s=0$. The average occupation number is
given by the sum of the residue of the pole and the cut
contribution. At long times the cut contribution vanishes at least as
a power law\cite{attanasio} and the asymptotic average occupation
number is given by  
\begin{eqnarray}
\left<n_{qk}(\infty)\right> &=& \frac{
n(0)+ \frac{1}{\pi  \Omega} \int^{\omega_c}_{\omega_{th}} d\omega ~
J(\omega) \left\{ 
\dfrac{1+N(\omega)}{\omega^2_+} + 
\frac{N(\omega)}{\omega^2_-} \right\}}{
1 - \frac{1}{\pi  \Omega}\int^{\omega_c}_{\omega_{th}} d\omega ~
J(\omega) \left\{\frac{1}{ 
\omega^2_+} - \frac{1}{\omega^2_-} \right\}}\; .\nonumber
\end{eqnarray}
The denominator of the above equation can be simplified considerably becoming simply $Z^{-2}$. The above equation is now written as
\begin{eqnarray}
\left<n_{qk}(\infty)\right> &=& {Z^2} \left[
n(0)+ \dfrac{1}{\pi  \Omega} \int^{\omega_c}_{\omega_{th}} d\omega ~  J(\omega) \left\{
\dfrac{1+N(\omega)}{\omega^2_+} + 
\dfrac{N(\omega)}{\omega^2_-} \right\} \right]\; .\nonumber
\end{eqnarray}
Comparing the above result with the one obtained exactly in the small coupling 
regime, \mbox{eq.(\ref{exres})}, we see that the two results coincide.

Obviously this quantum kinetic equation includes contributions from 
intermediate states that do not conserve energy and therefore provide
off-shell corrections. Although for the case in which the
quasiparticle pole is in the continuum, we see that asymptotically at
long times the distribution becomes similar to that obtained in the
Boltzmann approximation with the same relaxation 
rate, but at early times
the solution of the quantum kinetic equation differs appreciably from
the Boltzmann solution in that the relaxation 
rate vanishes at the initial time, whereas it is a constant for
Boltzmann.  The vanishing of the relaxation rate at the 
initial time is a consequence of the fact that the initial density
matrix is diagonal in the number representation, thus 
whereas the quantum kinetic equation describes correctly the initial
evolution, the Boltzmann equation has coarse grained 
over these time scales and misses the early time behavior.  

\subsection{Markovian approximation:}

If the particle occupation number varies on time scales larger than
the memory of the  
kernel in the kinetic equation, a Markovian approximation may be
reasonable. In the  
Markovian approximation, the particle occupation number $n(t')$ in
\mbox{eq.(\ref{kin})} 
is replaced by $n(t)$ and taken outside the integral.
This approximation would be justified in a weak coupling limit, in this
case when the spectral density of the bath $J(\omega)$ includes a small
coupling (as it will be specified in the next section) $\eta$. The
rational behind this approximation is the realization of multi-time
scales: a microscopic or short time scale given by $t \approx
1/\omega_p, \approx 1/\omega_{th}$ and another relaxation or long time
scale $t_1 \approx \eta t$.

Thus in the Markovian approximation, \mbox{eq.(\ref{kin})} becomes
\begin{eqnarray}
\left<\dot{n}(t)\right> &=&
\frac{1}{\pi  \Omega} \int d\omega J(\omega) \left\{
\dfrac{(1+N(\omega))\sin(\omega_+ t)}{\omega_+} + 
\dfrac{N(\omega)\sin(\omega_- t)}{\omega_-} \right\} \nonumber \\
& & \;+\; 
\frac{n(t)}{\pi  \Omega} \int d\omega J(\omega) \left\{
\dfrac{\sin(\omega_+ t)}{\omega_+} - 
\dfrac{\sin(\omega_- t)}{\omega_-} \right\}. 
\label{mark}
\end{eqnarray}

\noindent A computational advantage of this equation is that it
provides a {\em local} update procedure.  
A connection with the Boltzmann approximation is made with a second
stage of approximation, known in the Boltzmann literature as the
`completed 
collision approximation' and consists in taking the limit $t
\rightarrow \infty$ in the arguments of the sine functions in
eq.(\ref{mark}). Using the limiting distribution 
$$
{lim}_{t \rightarrow \infty} \frac{\sin[\omega_{\pm}t]}{\omega_{\pm}}
= \pi \delta(\omega_{\pm}) 
$$
which is used in the derivation of Fermi's Golden Rule, 
and noticing that only $\omega_-$ could vanish leads to the Boltzmann
expression
$$
\left<\dot{n}(t)\right> = \frac{J(\Omega)}{\Omega} \left[ N(\Omega)
-n(t) \right] 
$$
which coincides with the Boltzmann equation obtained above
[eq.(\ref{Boltzmann})] within the limit of validity of the perturbative
expansion since perturbatively $\Omega \approx \omega_0$ if $\Omega$
is taken as the position of the quasiparticle pole.  

\section{Numerical Analysis \label{secnum}}

In order to compare the particle number relaxation $n(t)$ between the
exact results \mbox{eq.(\ref{exactn})} and the various approximations
to the kinetic description \mbox{eq.(\ref{kin})}, Boltzmann and
Markovian, we 
have solved  numerically for a particular choice of the spectral density
of the bath.

We will choose the following model for $J(\omega)$
\begin{equation}
J(\omega)\,=\,
\eta\,\left(\omega\,-\omega_{th}\right)\,\theta(\omega-\omega_{th}) 
\,\theta(\omega_c-\omega).
\label{jspec}
\end{equation}
This is a generalization of the Ohmic bath in which  $J(\omega)$
vanishes for frequencies below a threshold frequency $\omega_{th}$ and
above a cutoff frequency $\omega_c$, and $\eta$ is a coupling
parameter. This is the simplest spectral density of the 
bath that allows us to model important features of a field theory and
illuminates the main aspects of  
 relaxational dynamics.

This form of the spectral density for the bath has been motivated by
previous studies of decoherence and dissipation in similar model 
theories\cite{ullersma}-\cite{beilok,unruh,yoshimura},
including model descriptions of entropy production and decoherence in
heavy ion collisions\cite{elze2}. It is the 
 simplest  realization
that allows us to vary parameters and investigate the different
regimes for the phenomena discussed in the previous  
section. By varying the coupling $\eta$ and the value of the bare (or
renormalized) frequency we can test the 
different scenarios. 

For the case of the quasiparticle pole embedded in the continuum the
 dimensionless 
 parameter that determines the separation of 
time scales is given for the spectral density (\ref{jspec}) by
$$
\frac{\Gamma}{\omega_p} \cong
\frac{\eta}{2\omega^2_p}(\omega_p-\omega_{th})\; . 
$$
\noindent When this ratio is $<<1$ the resonance is rather narrow and
there are many oscillations before the decay, the 
time scales are widely separated. In the other limit when this ratio
$\approx 1$ the particle is strongly coupled to the 
bath, resulting in a wide resonance and a potential for large
off-shell effects including effects related to the 
proximity of the peak of the resonance to the threshold.  

For $J(\omega)$ 
given by  eq.(\ref{jspec}), the real and imaginary parts
of $\tilde{\Sigma}_R(s=i\omega+\epsilon^+)$ are given by
\begin{eqnarray}
\Sigma_R(\omega) & = &  \frac{\eta }{\pi  }\left\{ \left( \omega_{th} +
           \omega \right) 
           \,\log \left[{\frac{\omega_{th}(\omega_c + \omega)}
            {\omega_c(\omega_{th} + \omega)}}\right] + 
        \left( \omega_{th} - \omega \right) \,
         \log \left|{\frac{\omega_{th}(\omega_c - \omega)}
             {\omega_c(\omega_{th} - \omega)}}\right|\, \right\}
           \nonumber \\ 
\Sigma_I(\omega)& = & {\eta}\,\left(\omega\,-\omega_{th}\right)\,\theta(
\omega -\omega_{th}) \,\theta(\omega_c-\omega)\; . \nonumber
\end{eqnarray}

The dynamical function $g(t)$ satisfies the equation of motion of the particle, eq.(\ref{xbar}).
In terms of the renormalized frequency $\omega_R$ given by
eq.(\ref{renofreq}), $g(t)$ can be shown to satisfy the following 
equation
$$
\ddot{g}(t) + \omega_R^2 g(t) + \frac{2}{\pi } \int_0^t dt' \int
d\omega \frac{J(\omega)}{ 
\omega} \cos\left[\omega(t-t')\right] \dot{g}(t') \,=\, 0 \; ; \;
g(0)=0 \; ; \; \dot{g}(0)=1. 
$$

We scale our results to an arbitrary unit of frequency and refer all
dimensionful quantities to this unit since the 
important physical quantities are dimensionless ratios (such as
$\omega/T$ etc).

Now we study different scenarios in detail.

\begin{figure}
\centerline{\epsfig{file=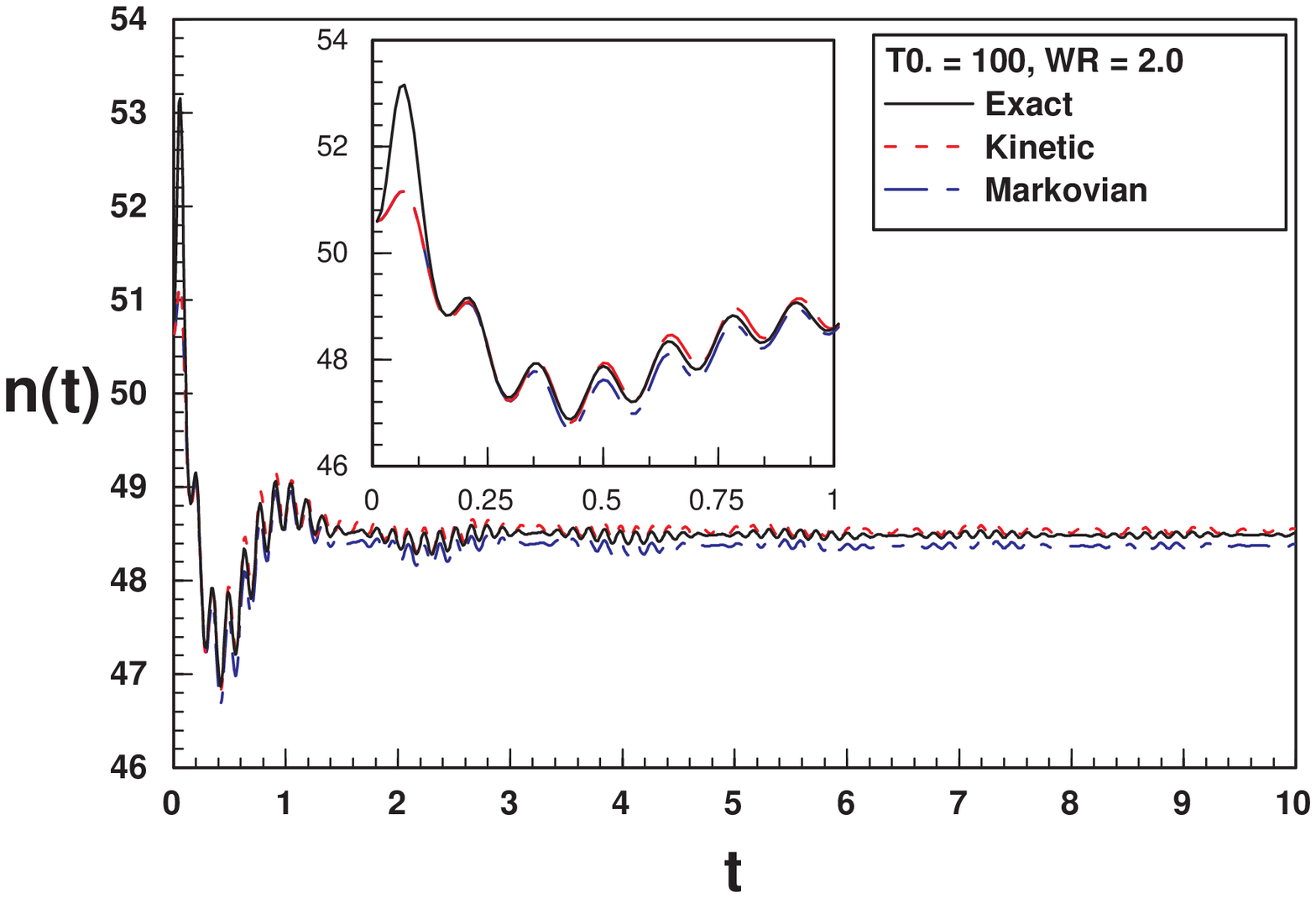,width=3.5in,height=3.5in}}
\centerline{ \epsfig{file=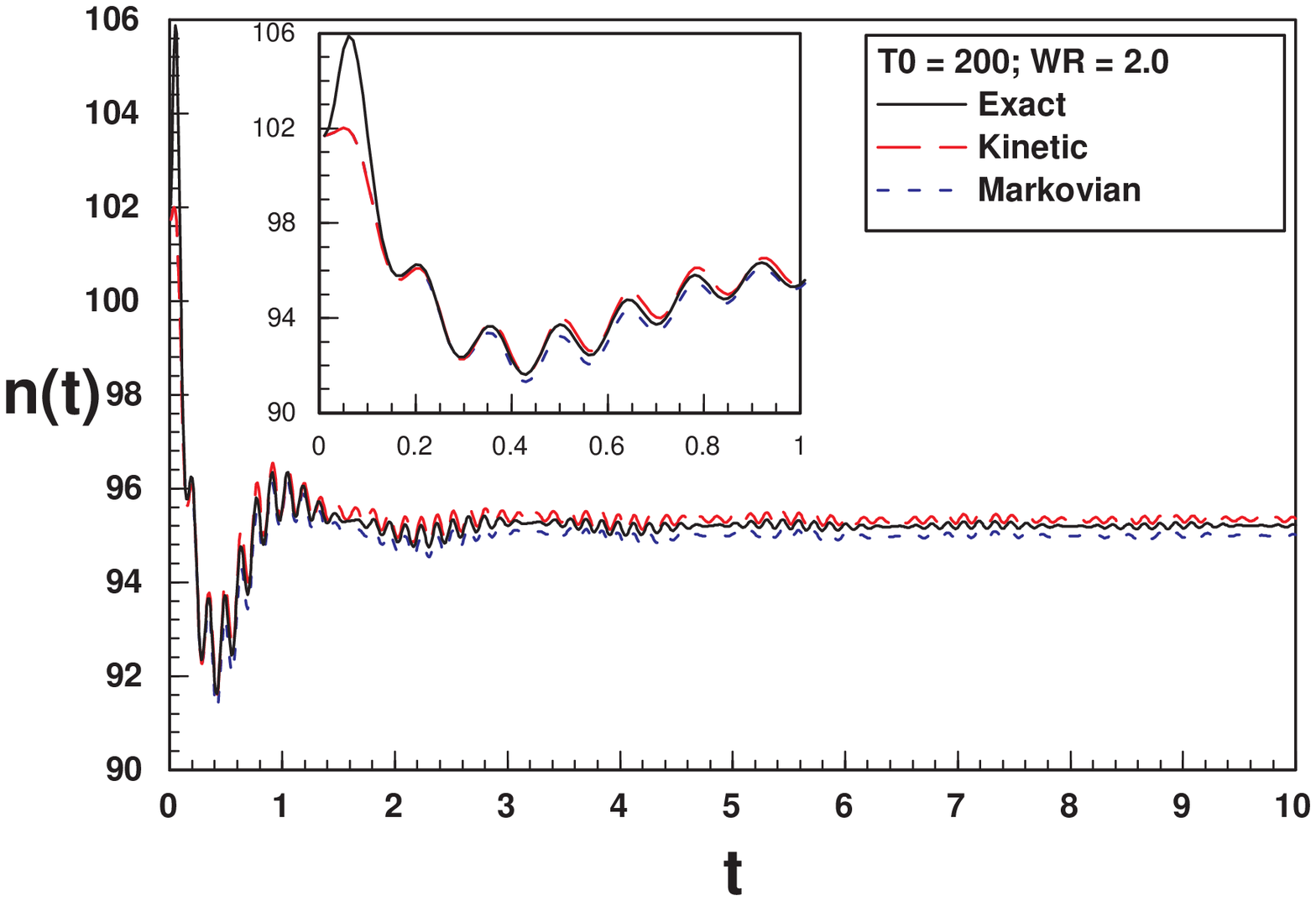,width=3.5in,height=3.5in} 
\epsfig{file=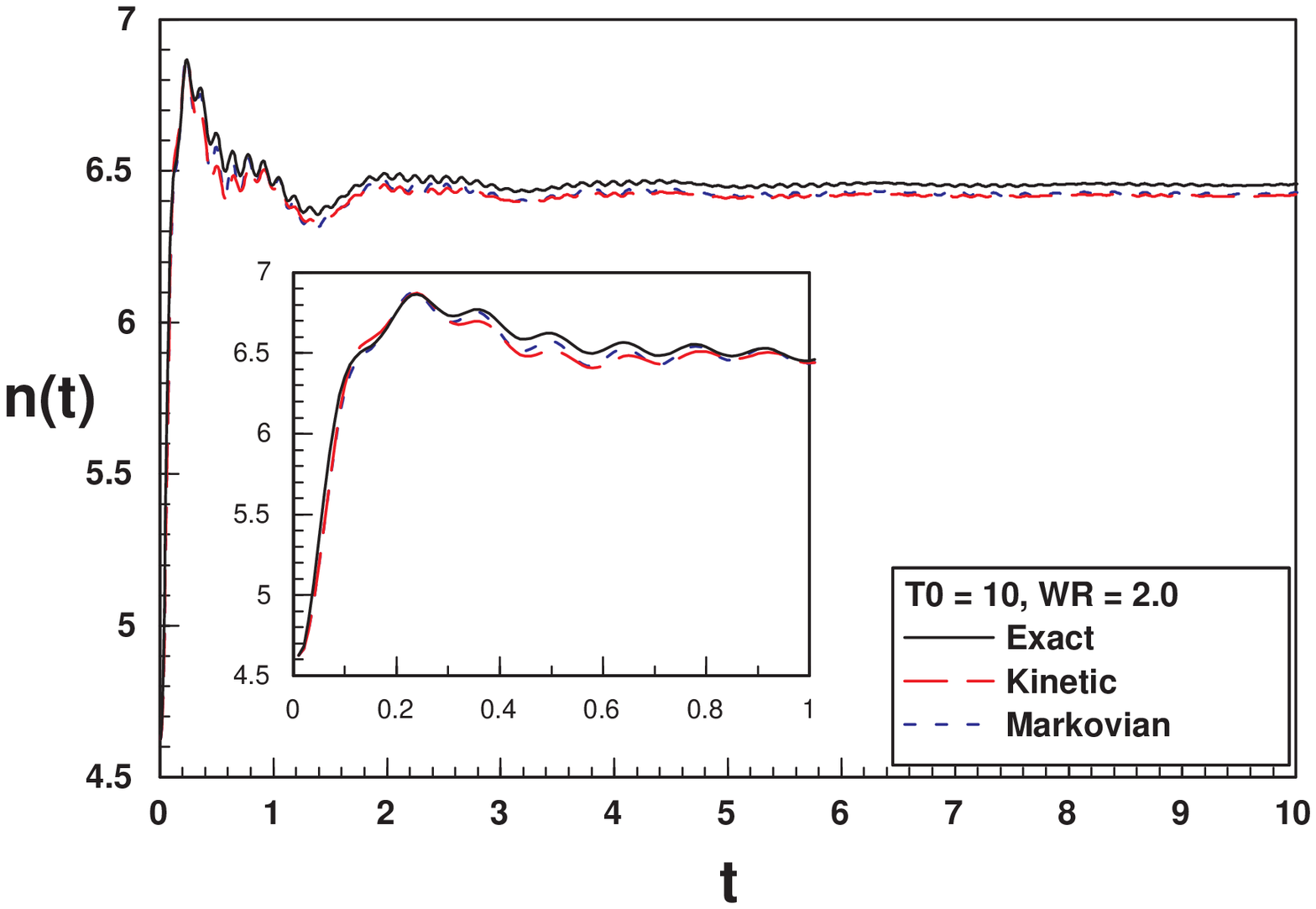,width=3.5in,height=3.5in}}
\caption{The expectation value of the particle occupation number given
by eq.(\ref{number}) for $\Omega=\omega_p$ for the case in which 
the pole is {\bf below} threshold for different particle temperatures 
$ T_0 = 100, 200 $ and $ 10 $ and bath temperature $T =  100$. 
The pole is at $\omega_p = 1.95719 $ and $Z = 0.95621 $.
The numerical parameters are $ \eta = 0.85 $,
$\omega_c = 45, \; \omega_{th} =  5$ and $\omega_R = 2 $. \label{neb85}}
\end{figure}

Figure \ref{neb85} shows the case for which the dressed particle pole
is below threshold. In this case the Boltzmann 
equation predicts that no relaxation occurs because the imaginary part
of the self-energy evaluated on shell (damping 
rate) vanishes. The exact solution, and the quantum kinetic
approximation along with the Markovian limit all predict 
non-trivial relaxation, and for this weak coupling case all agree to
within few percent. Obviously in this case the 
relaxation is solely due to off-shell effects since the dissipative
effects associated with processes that conserve 
energy (on-shell) vanish. The inset of the figures shows the dynamics
of dressing of the particle and the time 
scales predicted by the exact result are well reproduced by both the
quantum kinetic equation and its Markovian approximation. 
 
\begin{figure}
\centerline{\epsfig{file=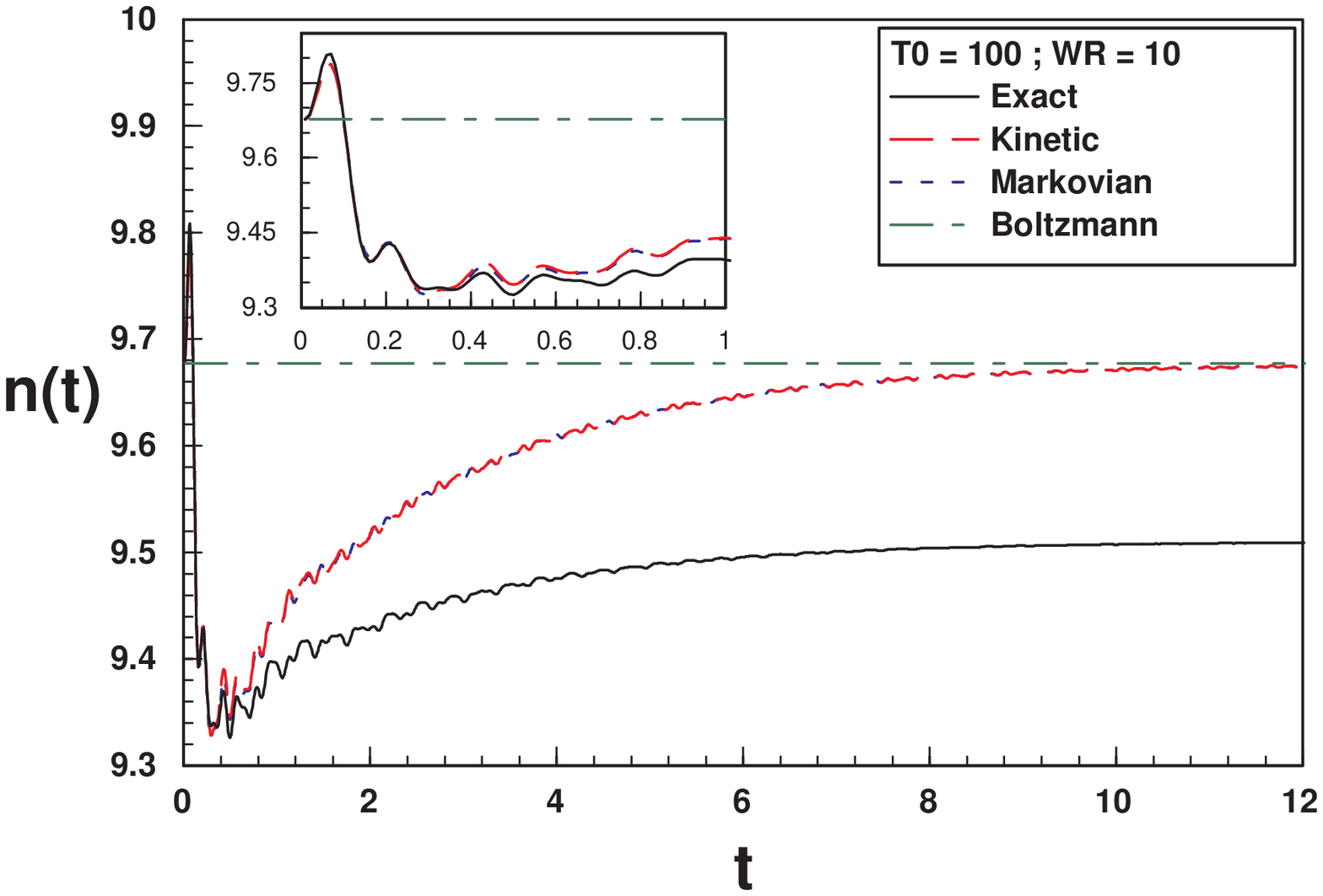,width=3.5in,height=3.5in}}
\centerline{ \epsfig{file=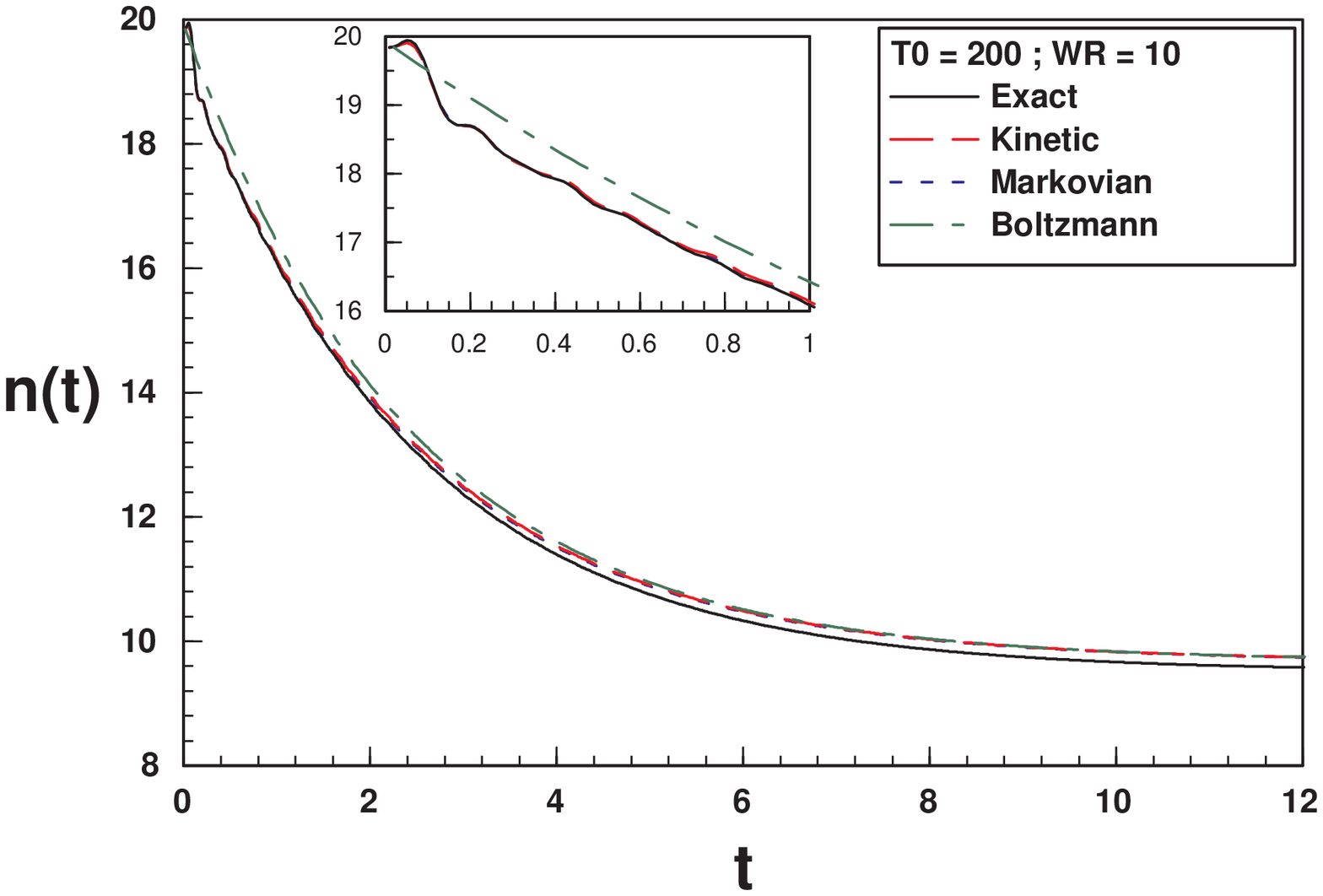,width=3.5in,height=3.5in} 
\epsfig{file=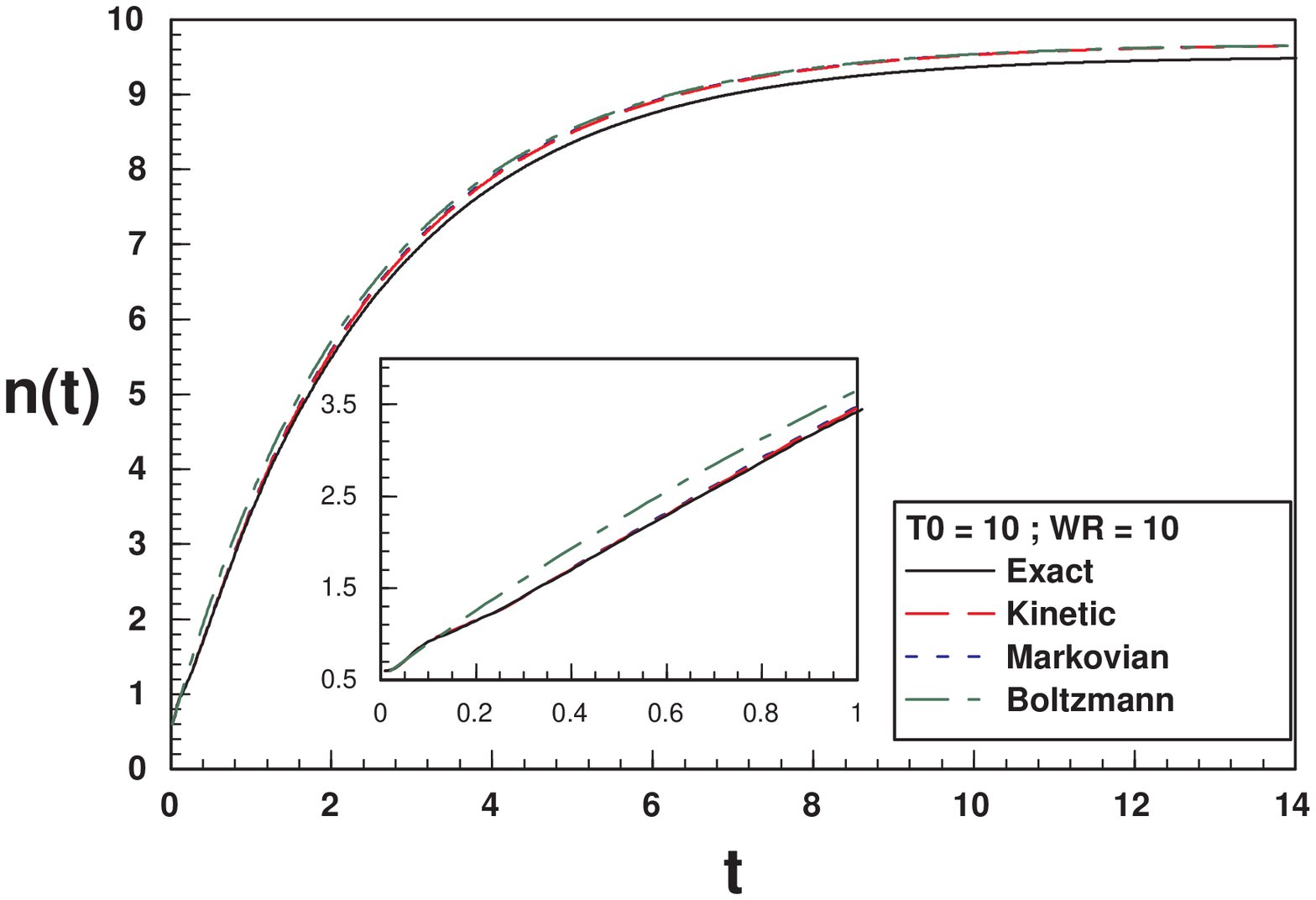,width=3.5in,height=3.5in}}
\caption{The expectation value of the particle occupation number given
by eq.(\ref{number}) for $\Omega=\omega_p$ for the case in which 
the pole is {\bf above} threshold for different particle temperatures 
$ T_0 = 100, 200 $ and $ 10 $ and bath temperature $T =  100$. 
The pole is at $\omega_p = 9.83397 $ and $Z = 0.99631$.
The numerical parameters are $\eta = 0.85$,
$\omega_c = 45, \; \omega_{th} = 5 $ and $ \omega_R = 10$, resulting in
$\Gamma / \omega_p \approx 0.02$. \label{ne85}}
\end{figure}

In contrast, fig.\ref{ne85} shows the case for which the
quasiparticle pole is in the continuum but with 
a narrow width  $\Gamma / \omega_p \approx 0.02$. The bath temperature
is fixed $T=100$ and  
the initial temperature of the particle ($T_0$), 
is varied in a wide range. We
notice that in the case in which the temperature 
of the bath and that of the bare particle are the same, the Boltzmann
equation predicts no relaxation because the gain and 
loss processes balance exactly, this is the straight line in the graph
for bare particle temperature $T_0 = 100$. The exact solution  
as well as  the kinetic and Markovian approximation predict
relaxation, the kinetic and the Markovian approximations are very close to the 
exact expression. Analytically we know that the exact, Markovian 
and kinetic will asymptotically approach the Boltzmann result (with
very small corrections)  in this
very narrow width case. Obviously the time scales for relaxation
and the early time dynamics are features not reproduced by the
Boltzmann equation and clearly a result of off shell  effects, since
all of the energy conserving detailed balance processes are
contemplated by the Boltzmann equation.  

\begin{figure}
\centerline{ \epsfig{file=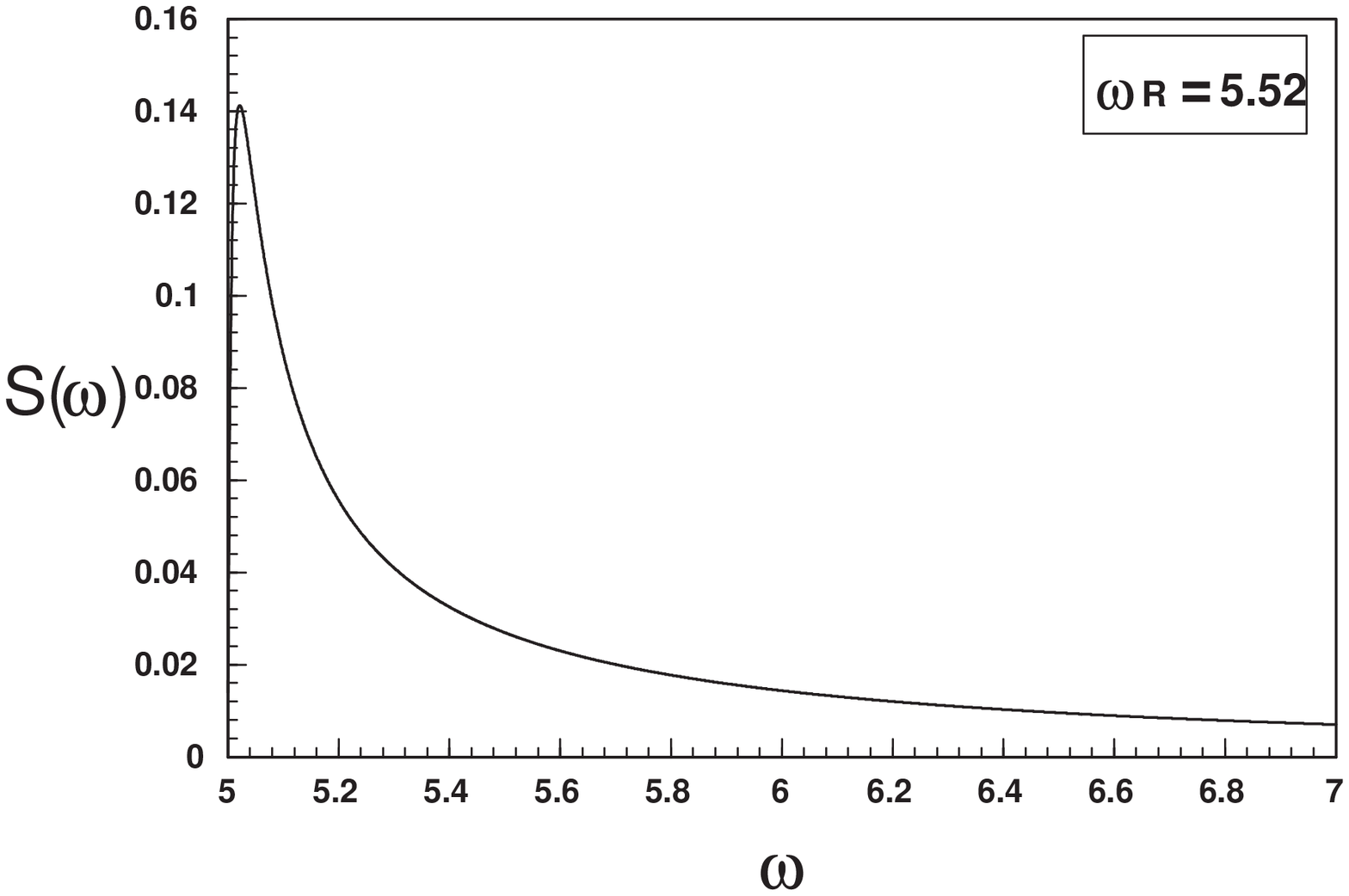,width=3.25in,height=2.75in}
\epsfig{file=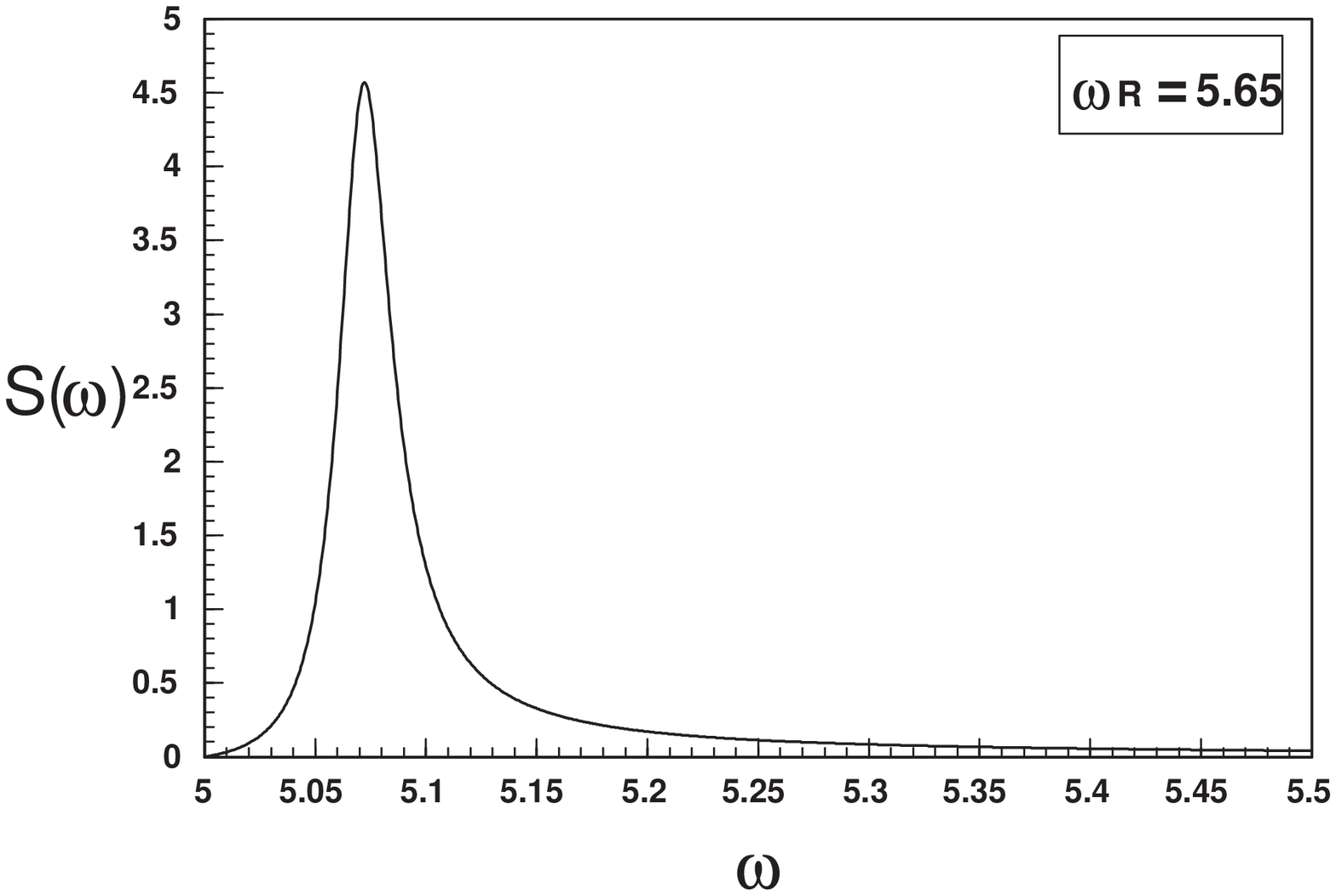,width=3.25in,height=2.75in}}
\centerline{ \epsfig{file=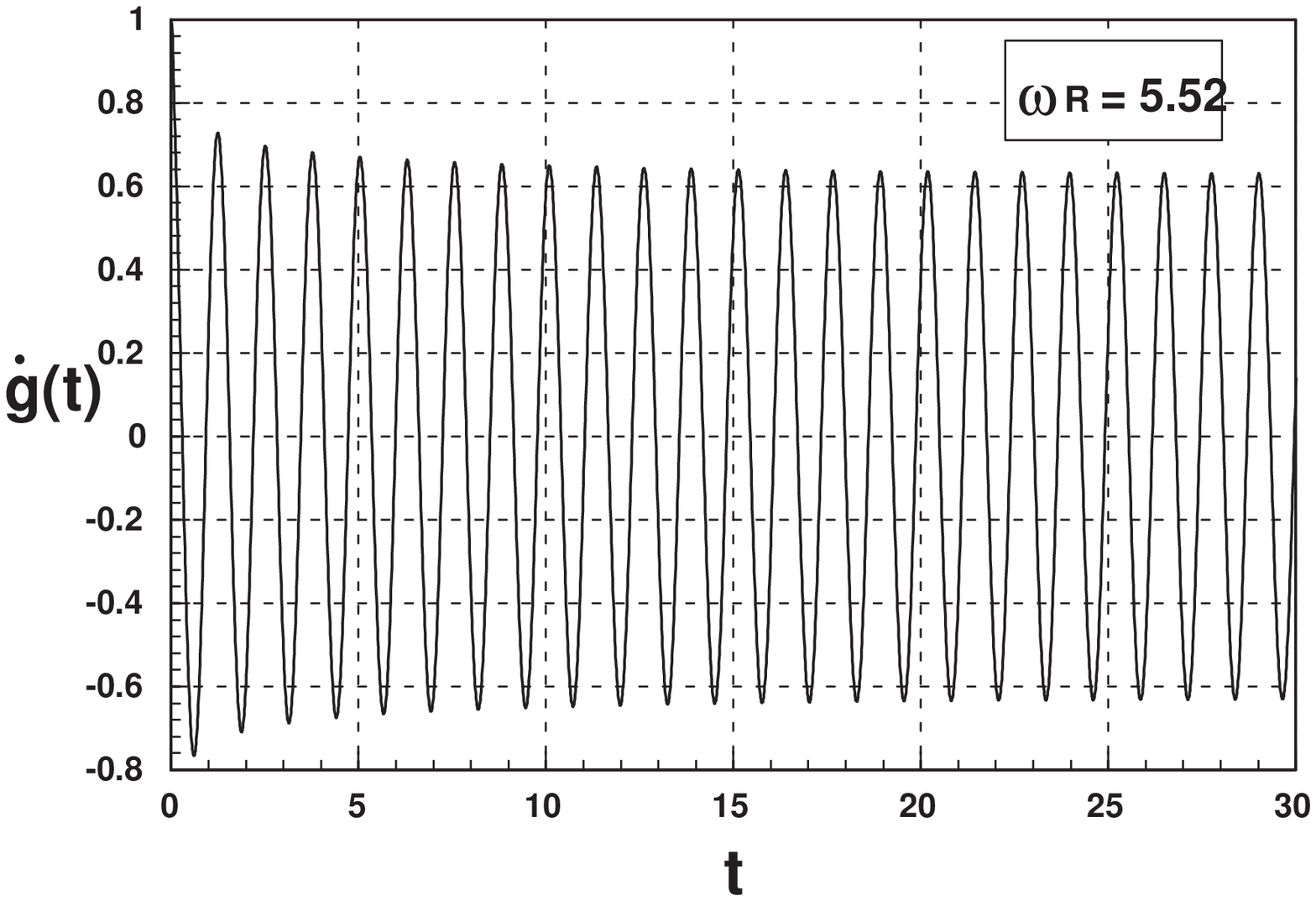,width=3.25in,height=2.75in} 
\epsfig{file=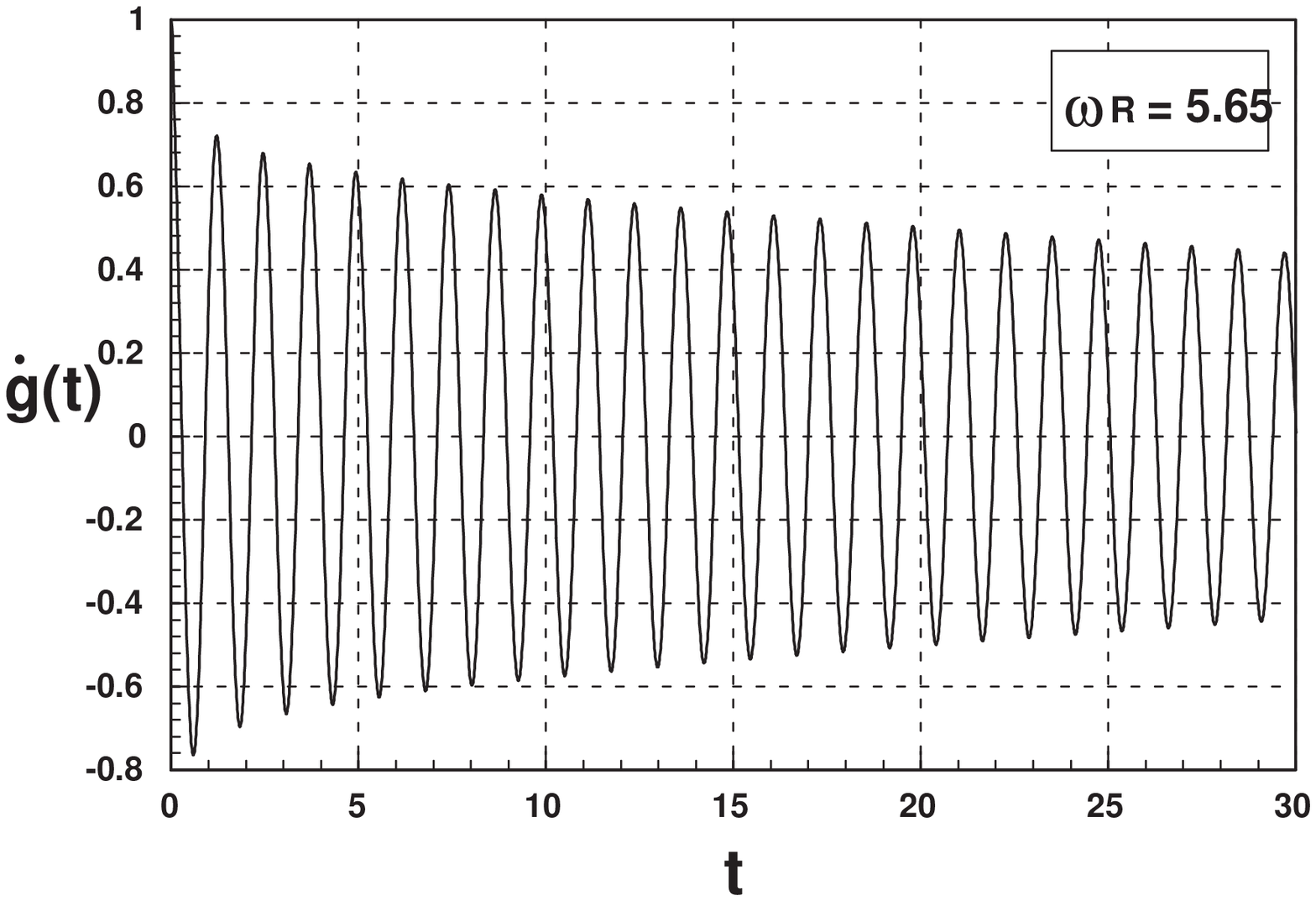,width=3.25in,height=2.75in}}
\caption{The functions $S(\omega)$ and  $\dot{g}(t)$  for the cases in
which the pole 
is {\bf just below} (left column) and just above (right column) the
threshold frequency  
($\omega_{th} =$ 5). 
In the left column,  $\omega_R = 5.52 , \; \omega_p = 4.98083 $ and
$Z = 0.63845 $ while in the
right column $\omega_R = 5.65, \; \omega_p = 5.07373 $ and $Z = 0.69959$.
The numerical parameters are $\eta = 3.0 $  and $ \omega_c = 55 $, with
$ \Gamma / \omega_p \approx 0.005 $. \label{nearth}} 
\end{figure}

\begin{figure}
\centerline{ \epsfig{file=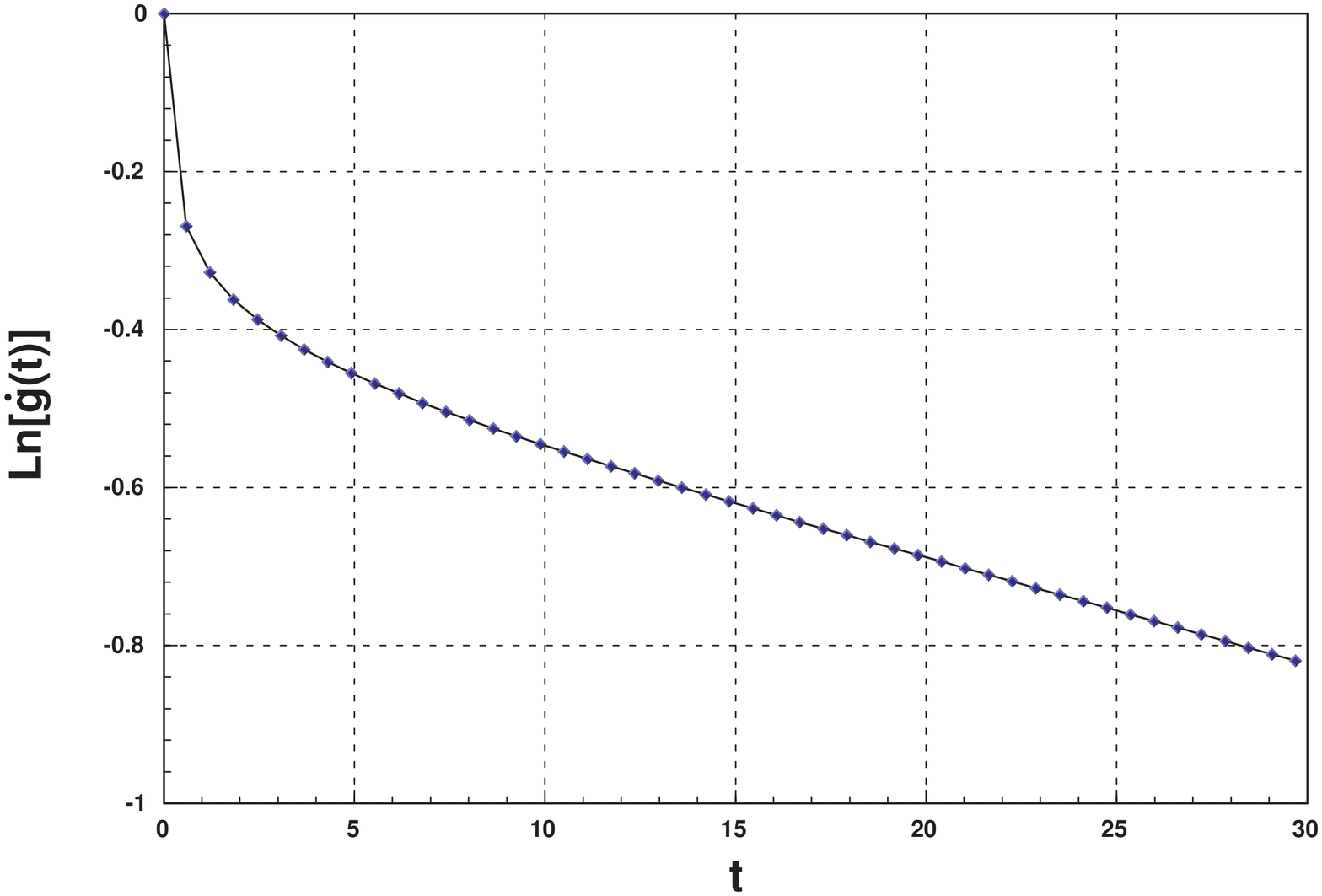,width=3.75in,height=3.75in}}
\caption{The  logarithm of the maxima of $\dot{g}(t)$ vs $t$ for $\eta
=$ 3.0 and $\omega_c = 55, \; \omega_{th}=5 $,
$\omega_R = 5.52, \; \omega_p = 4.98083 $ and 
$Z = 0.63845 $, corresponding to the right column of fig.\ref{nearth}. 
\label{quasiform}}
\end{figure}

\begin{figure}
\centerline{ \epsfig{file=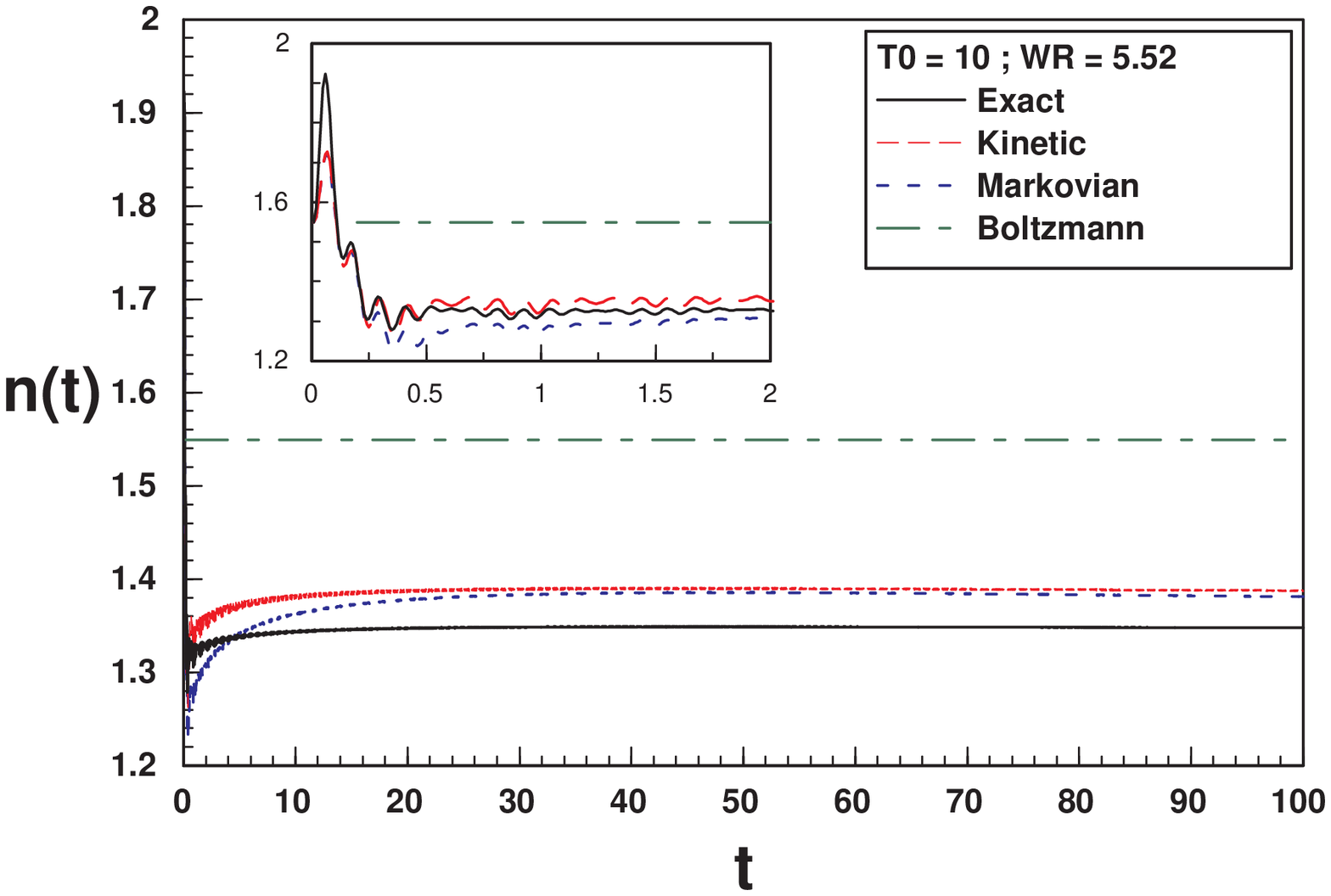,width=3.25in,height=2.75in}
\epsfig{file=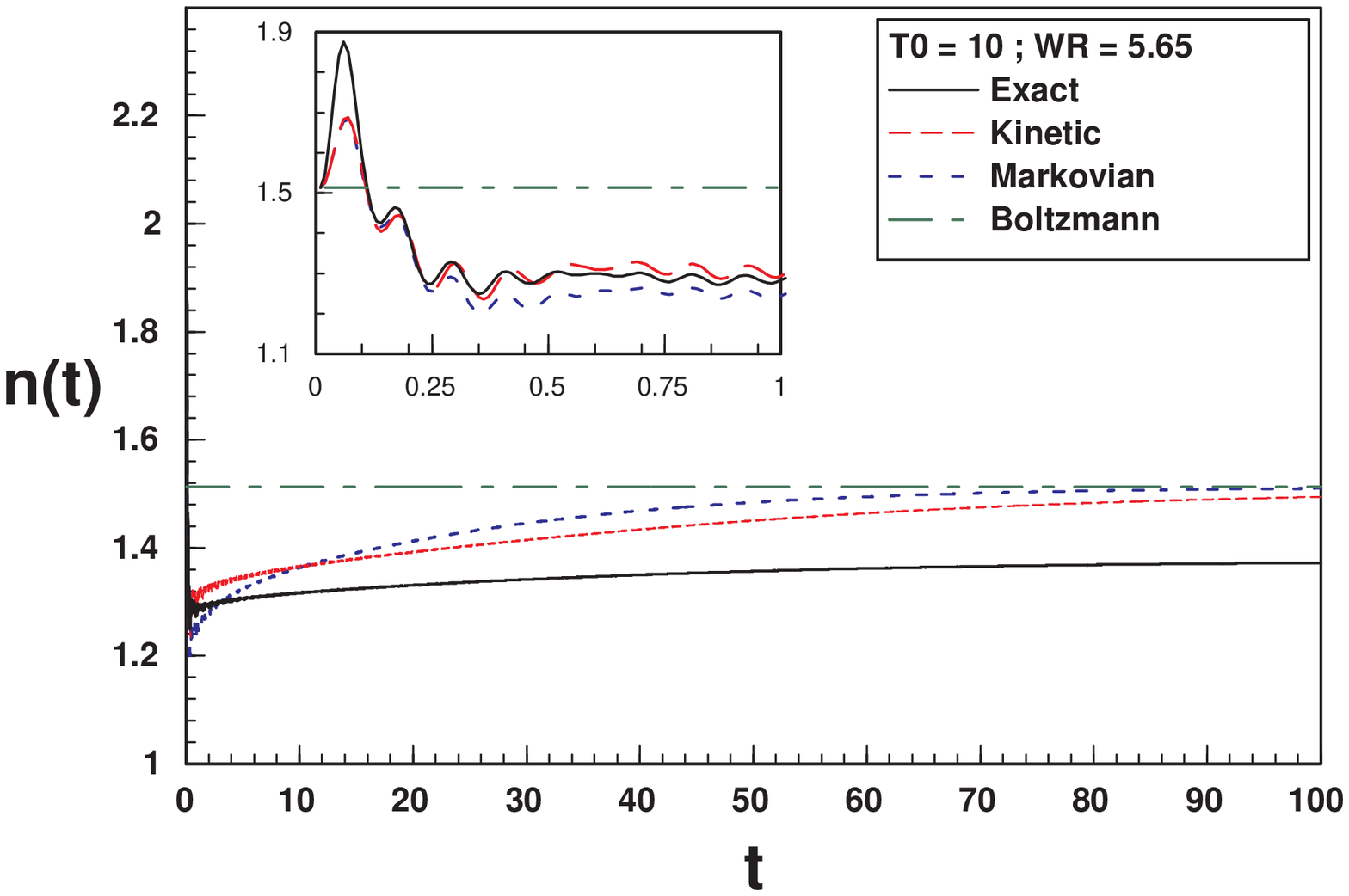,width=3.25in,height=2.75in}}
\caption{ The expectation value of the number operator
eq.(\ref{number}) for the same values of 
the parameters as in figure(\ref{nearth})(left figure corresponds to
pole below threshold and right figure to 
the pole above threshold) for the case of equal particle and bath
temperature $ T_0=T=10 $. The insert shows the 
early time behavior. \label{nbeab}}
\end{figure}

Figure \ref{nearth} compares two situations: the left figures
correspond to the case of a  pole just 
slightly below (dressed particle) and the right figures just slightly
above  threshold (quasiparticle). 
This case provides for strong renormalization 
effects because the wave function renormalization  departs
significantly from one. The left figure for $\dot{g}(t)$ depicts 
clearly the dressing time of the particle, with $\dot{g}(0)=1$
we see that after a short time the asymptotic value 
$ \dot{g}(t) \approx Z \cos(\omega_p t) $  is achieved. This figure thus
reveals {\em two} time scales, one associated with 
the oscillation scale of the dressed particle $ 1/ \omega_p$ and the
other associated with the decay to the asymptotic form, 
this time scale determines the dressing time of the particle and
for the case under consideration corresponds to  
just a few oscillations. This dressing time scale clearly depends on
the details of the spectral density since it determines the early time
dynamics after the preparation of the initial state. The right figure
for $\dot{g}(t)$ presents 
{\em three different} time scales: initially there is the time scale
of formation of the quasiparticle, very similar 
to the left figure, the time scale associated with the quasiparticle
pole $\approx 1/\omega_p$ and finally the time 
scale associated with the exponential decay. The formation time scale
and that of exponential decay can only be resolved 
in the narrow width approximation, in this particular example $\Gamma
/ \omega_p \approx 0.005$ and the time scales associated with the
quasiparticle formation from the initial state and 
exponential relaxation can be resolved. These are clearly displayed in
fig.\ref{quasiform} where the logarithm of the maxima of
$\dot{g}(t)$ is plotted versus time. In fig.\ref{nbeab} we show  
the expectation value of the number operator eq.(\ref{number}) for
$\Omega=\omega_p$ for the same values of 
the parameters as in fig.\ref{nearth} (left figure corresponds to
pole below threshold and right figure to 
the pole above threshold) and equal particle and bath temperature
$ T_0 = T = 10 $. Whereas the Boltzmann equation predicts 
again no relaxation, in the left figure  because the damping rate
vanishes and in the right figure because 
 the on-shell gain and loss processes balance each other, the exact
and quantum kinetics 
description of relaxation both predict non trivial evolution of the
dressed particle and quasiparticle distribution functions
respectively. The left figure  
shows that whereas the quantum kinetic and Markovian evolution are not
too different from the exact, asymptotically all of them  depart
significantly from Boltzmann. The early time dynamics predicted by the
Markovian and quantum kinetics  are 
very close to the exact expression. In the right figure, corresponding
to a narrow resonance we see that asymptotically 
the quantum kinetic and Markovian evolution asymptotically approach
the Boltzmann result but obviously the early and 
intermediate time dynamics is remarkably different. Furthermore the
exact result reaches an asymptotic value that is 
very different from Boltzmann, as a result of the strong quasiparticle
renormalization effects, with $Z$ departing  
significantly from one [see fig. \ref{nearth}]. Despite the fact
that the resonance is rather narrow, its proximity to threshold results in 
strong off-shell effects. 

\begin{figure}
\centerline{ \epsfig{file=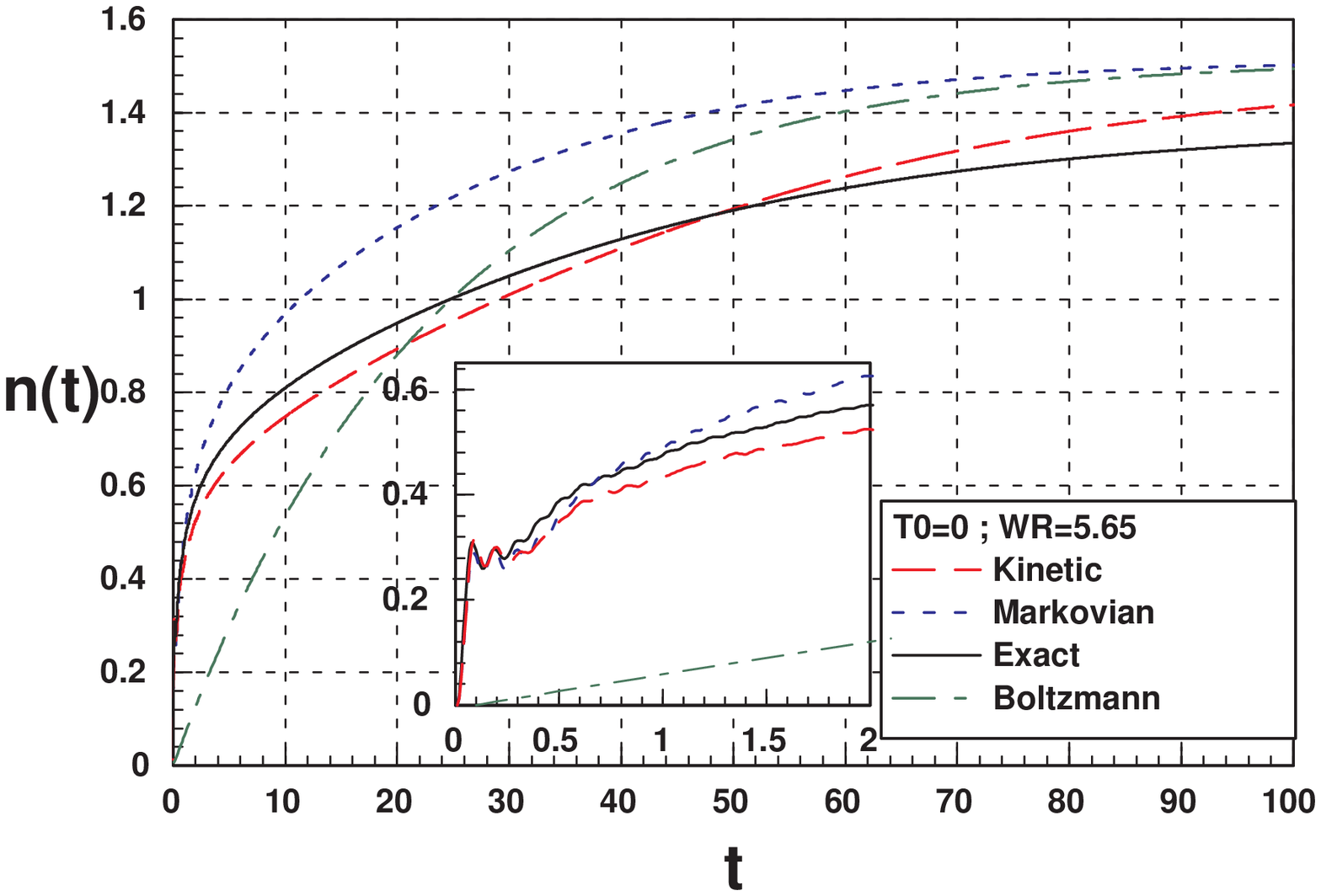,width=3.25in,height=2.75in} 
\epsfig{file=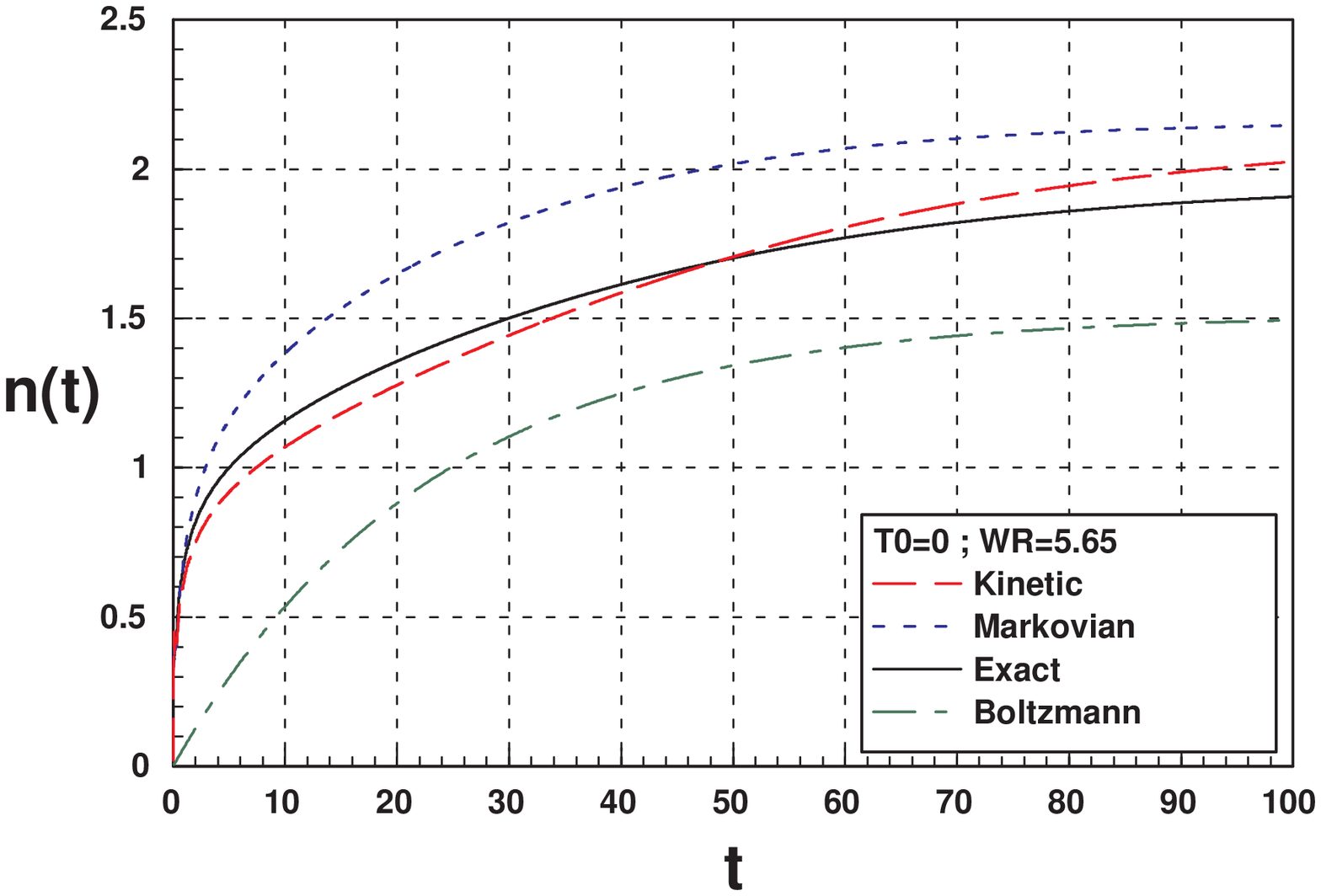,width=3.25in,height=2.75in}}
\caption{The expectation value of the number operator
eq.(\ref{number}) for the same values of 
the parameters as for the right column in
figure(\ref{nearth})(quasiparticle pole above threshold), $\eta = 3.0 $
and $\omega_c = 55, \; \omega_{th}=5 $, 
$\omega_R = 5.65, \; \omega_p = 5.07373 $ and $Z = 0.69959 $.
The temperature of the bath is $T=10$ and
zero initial particle temperature ($ T_0=0 $), the exact,
Markovian and kinetic curves have been divided by the wave function
renormalization $Z$ in the rightmost figure. The insert in the left
figure shows the early time behavior.\label{nearth2}} 
\end{figure}

Figure \ref{nearth2} is perhaps one of the
most illuminating. The parameters are the same 
as for the right part of fig.\ref{nearth}, i.e. with the
quasiparticle pole in the continuum and close to threshold, 
the bath temperature is $T=10$ and the particle is initially at zero
temperature. In the left figure we plot the Boltzmann, 
exact, quantum kinetic and Markovian evolutions respectively for the
expectation value of the number operator (\ref{number}) for $\Omega =
\omega_p$, whereas the right figure corresponds to dividing by $ Z $ the
results of the exact, quantum kinetics 
and Markovian evolutions to make contact with the quasiparticle number
operator (\ref{quasi}). This figure clearly shows that the Boltzmann
approximation coarse grains over the 
early time behavior and completely misses the formation time scales
and the early details of relaxation.

\begin{figure}
\centerline{ \epsfig{file=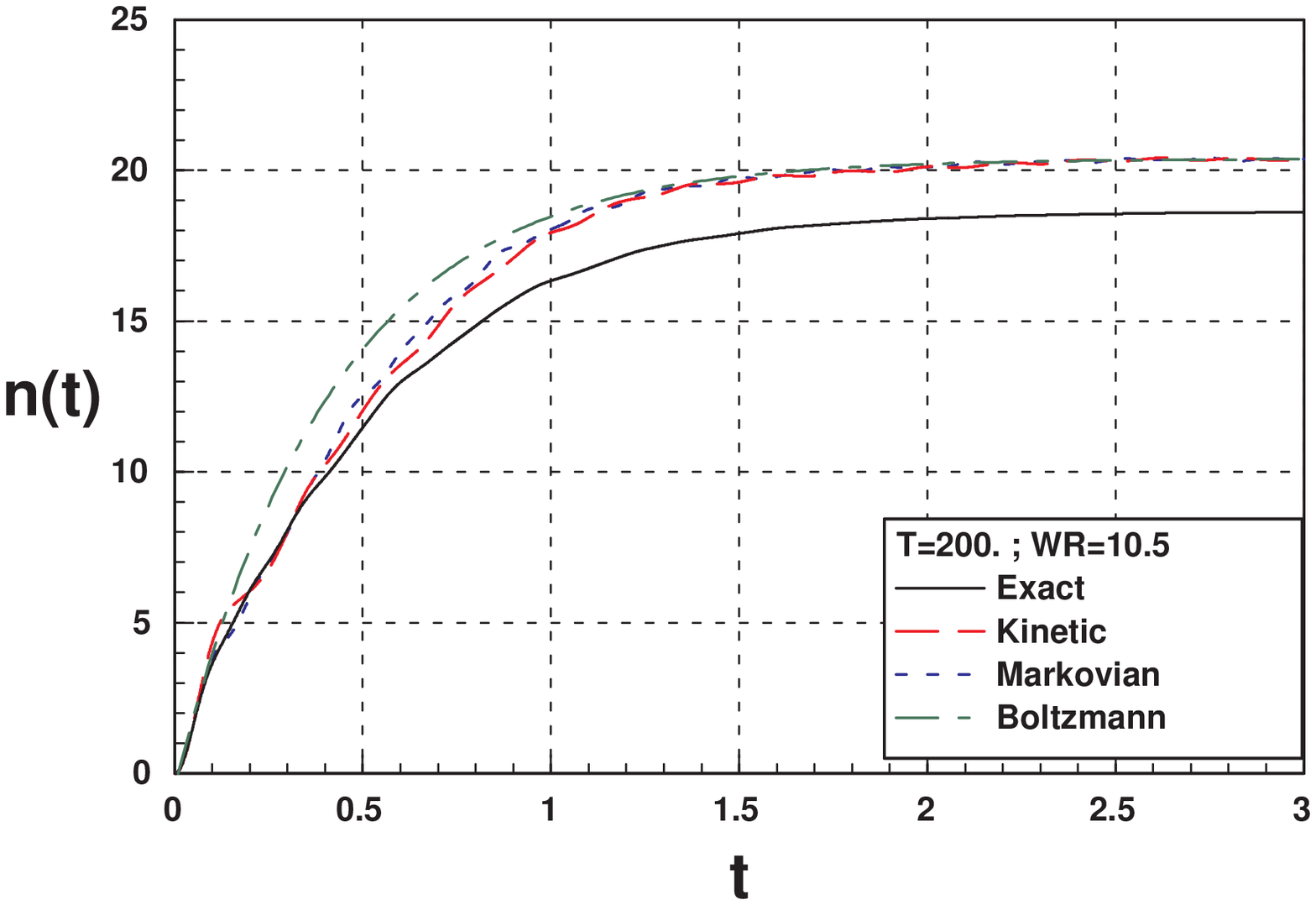,width=3.75in,height=3.75in}}
\caption{The expectation value of the number operator
 eq.(\ref{number}) for  $\eta = 5.0 $ and $\omega_c = 40 $,
 $\omega_{th}=5, \; \omega_p = 9.58, \; Z = 0.982 $ and bath
temperature $ T = 200 $, corresponding to $ \Gamma / \omega_p \approx 0.1 $.   
\label{rholike}}
\end{figure}

Finally, fig.\ref{rholike} presents the evolution of the
quasiparticle distribution for a case of a strong coupling 
regime resulting in a wide resonance: $\eta = 5\; ; \;
\omega_{th}=5.0\; ; \; \omega_c=40 \; ; \; \omega_p=9.58\; ; \;  
Z=0.982$ for a bath temperature $T=200$ and zero initial particle
temperature. The ratio $\Gamma / \omega_p \approx 0.1$ is 
comparable to that of a realistic vector meson, such as for example
the neutral $\rho$  meson. The inset in the figure displays the early
time behavior.  
 We see in this figure that whereas the early 
time behavior is similar for the exact and approximate evolutions, which
is a consequence of zero initial temperature for the particle, at times
of the order of the relaxation time there is a dramatic
departure. Furthermore the Boltzmann approximation predicts a very
different early time 
evolution because it coarse grains over the formation time of the
quasiparticle.   

Whereas the quantum kinetic evolution and its Markovian
approximation track very closely the Boltzmann, the exact evolution is
approximately 
$15\%$ smaller resulting in a smaller population of resonances
asymptotically. The departures in the exact result are a consequence 
of off-shell effects associated with a large width of the resonance,
since in the narrow resonance approximation and for $Z$ so close to
one, the asymptotic limit of the exact solution coincides with that of
the 
Boltzmann equation.

\section{Conclusions and implications \label{seccon}}

The goal of this article is to study the dynamics of thermalization
including off-shell effects that are not incorporated in a Boltzmann 
description of kinetics. In particular the focus is to assess the
validity of the Boltzmann approximation as well 
as non-Markovian and Markovian quantum kinetic descriptions of
relaxation and thermalization in a  
model that allows an exact treatment.

Although the model treated in this article allows an exact solution
and therefore provides an arena to test the 
regime of validity of several approximate descriptions of kinetics and
compare to an exact result, it is obviously 
not a full quantum field theory. Specific field theory models used to
obtain a microscopic description of thermalization 
and relaxation will certainly contain details that are not  captured
by the model investigated here. However, from the 
exhaustive analysis in this article we believe that some of the
results obtained 
here are fairly robust and trascend any particular model. These are
the following: 

\begin{itemize}

\item{{\bf Boltzmann vs. quantum kinetics:} A necessary criterion for
the validity of a Boltzmann approach is that there is a clear and wide
separation of time scales between 
the microscopic time scales and the time scales of relaxation. This is
typically the situation in which quasiparticles 
correspond to very narrow resonances in the spectral functions and
their lifetime is much longer than the typical microscopic scales. If
perturbation theory is applicable {\em and} the quasiparticle
resonance is narrow and its position is far away from thresholds, then
a Boltzmann description is likely to be reliable for time scales longer than the formation time
of the quasiparticle. When there is
competition of time  
scales or the early stages are physically relevant a full quantum kinetic equation must be obtained. }

\item {{\bf Microscopic time scales:} In order to determine the
microscopic time scales the first step is to determine 
the position of the resonances or quasiparticles, i.e. the
quasiparticle pole including  the medium effects. The bare
particle poles  {\em do not} determine the microscopic time
scales. Obviously for weakly interacting theories the position 
of the bare and quasiparticle poles will be very close and the
microscopic time scales are similar.} 

\item{{\bf Relaxational time scales:} An estimate of the relaxational
time scale is determined by the width of the 
resonance, $\Gamma$, a wide separation of time scales that would
provide a necessary condition for the validity of  
a Boltzmann approximation would require that $\Gamma / \omega_p <<1$.} 

\item{{\bf Thresholds:} Although  a wide separation of time scales is
a necessary condition for the validity of a 
Boltzmann approach, it is not sufficient. In particular when the
position of the resonance is too close to threshold, 
there will be important corrections to the long and short time
dynamics arising from the behavior of the spectral  
density at threshold. Threshold effects can lead to strong
renormalization of the amplitude of the quasiparticle pole (wave 
function renormalization) that results in sizable distortions of the
equilibrium distributions as compared to 
the free particle ones. In particular we have seen how thermalization
is achieved but with large corrections in the 
quasiparticle distribution functions from the usual Bose-Einstein form.}

\item{{\bf Formation times, Markovian vs. non-Markovian kinetics:} The
model that we have studied allowed us to explore 
the concept of the formation time of a quasiparticle. This concept is
simply unavailable within a Boltzmann approach, since 
the Boltzmann equation coarse grains over the formation time
scales. This is clearly revealed in figures \ref{quasiform}
and \ref{nearth2}. Both the non-Markovian quantum kinetics and its
Markovian approximation include off-shell 
effects and capture the early time dynamics associated with the
formation of the quasiparticle. The formation time  
of a quasiparticle becomes relevant if the initial state is very far
from equilibrium, since in a non-linear evolution, 
large initial departures can result in large corrections in the
asymptotic region. An important message learned in 
this work is that even in strongly coupled cases a non-Markovian
quantum kinetic description provides a very good 
approximation to the correct dynamics for most of the relevant time
scale. A Markovian approximation that is obtained 
by extracting the distribution functions from inside the non-local
time integrals, but without taking the interval of 
time to infinity offers a viable description, which is close to the
exact evolution and that of the non-Markovian quantum kinetics at weak
and intermediate 
couplings. Its main advantage is computational because this approximation provides a local
update equation.  }

\item{{\bf Implications for Field Theory:}
The lessons learned in this article allow us to provide some sound
implications for realistic field theoretical models. 
In particular for example in the dynamics of thermalization of the
quark-gluon plasma, we are led to the conclusion 
that a Boltzmann description of thermalization could be reliable for
{\em hard} partons for which i) the perturbative 
(medium corrected) scattering cross sections and ii) the initial
parton distribution seem to lead to a separation between the (short)
microscopic scale (determined by the inverse of the hard momentum of
the parton) and the relaxation time scale 
for partons with typical energies larger than say about 1 Gev\cite{geiger}. 

For soft degrees of freedom we envisage strong departures from a
Boltzmann description and the necessity of a quantum 
kinetic description. This view is supported by  recent studies of
anomalous relaxation of quasiparticles in ultrarelativistic
plasmas\cite{iancu,boyprem,boyphoton} in which infrared and threshold
effects lead to very different relaxational dynamics as compared to
the predictions based on quasiparticle resonances\cite{iancu}.  
In this case, as advocated in ref.\cite{boyprem}
a quantum kinetic description begins by resumming the perturbative
expansion for the time evolution of the distribution 
function for the quasiparticles and keeping memory effects in a full
non-Markovian description. A Markovian approximation 
that leads to local update equations can be obtained in weak coupling
but without invoking energy conservation in 
intermediate states. This Markovian approximation has been shown in
this article to be very close to the non-Markovian 
quantum kinetics. Unlike a fully coarse grained Boltzmann description,
this intermediate Markovian equation describes 
rather well the initial stages of quasiparticle formation and offers
a local update alternative to non-local kinetics which includes 
off-shell effects.

We expect large corrections to Boltzmann relaxation in the case of
wide hadronic resonances such as for example  
vector mesons near the chiral phase transition. For typical vector
mesons both effects that lead to strong corrections 
are present: the ratio of the width of the resonance to the position
of the resonance (quasiparticle pole) is not 
much smaller than one, for example in the case of the $\rho$ vector
meson is about $0.2$ in vacuum, furthermore in medium, the position of the resonance 
is not too far from thresholds. In these cases we conclude that a non-Markovian quantum kinetic description will
be needed to understand reliably the abundance of resonances in the medium and their thermalization time
scales. Departures from Boltzman abundances in medium could lead to experimentally observable effects in 
dilepton and photon spectra from hadronic resonances for example. 

As presented in this article and advanced in previous work\cite{boyprem,boyphoton}, a non-Markovian, quantum kinetic description can be provided from a microscopic field
theory model by beginning with perturbation theory, and resumming the perturbative expansion including off shell
effects. The details of this program in a full quantum field theory applied to the case of resonance relaxation will be provided in future work.}

\end{itemize}

\section{Acknowledgements}  D. B. thanks the
N.S.F for partial support through the grant award: PHY-9605186 the
Pittsburgh Supercomputer Center for  
grant award No: PHY950011P and LPTHE for warm hospitality.  
S. M. A. thanks King Fahad University of Petroleum and Minerals (Saudi Arabia) 
for financial support.
R. H. was
supported by 
DOE grant DE-FG02-91-ER40682.  The authors  acknowledge partial support by
NATO.

\newpage



\newpage



%
%

%
%

%
%

\end{document}